\documentclass[aps,prd,reprint,amsmath,amssymb,superscriptaddress,longbibliography,showkeys,fleqn]{revtex4-2}

\usepackage{graphicx}
\usepackage{dcolumn}
\usepackage{bm}
\usepackage{etoolbox}
\usepackage{silence}
\usepackage{hyperref}
\usepackage{orcidlink}
\usepackage{microtype}

\csletcs{bibdataoutaps}{@bibdataout@aps}
\patchcmd{\bibdataoutaps}{author="08"}{author="48"}{}{}
\csletcs{@bibdataout@aps}{bibdataoutaps}
\patchcmd{\subequations}
{\theparentequation\alph{equation}}
{\theparentequation.\alph{equation}}{}{}
\AtBeginEnvironment{thebibliography}{\sloppy\hbadness=2000\relax}

\begin{document}

\title{Curvature-driven spin transport in rank-two string-Carroll hydrodynamics}

\author{Nikko John Leo S. Lobos \orcidlink{0000-0001-6976-8462}}
\email{nslobos@ust.edu.ph}
\affiliation{Electronics Engineering Department, University of Santo Tomas, Espa\~na Boulevard, Sampaloc, Manila 1008, Philippines}

\author{Reggie C. Pantig \orcidlink{0000-0002-3101-8591}}
\email{rcpantig@mapua.edu.ph}
\affiliation{Physics Department, School of Foundational Studies and Education, Map\'ua University, 658 Muralla St., Intramuros, Manila 1002, Philippines}

\date{\today}

\begin{abstract}
 We derive the rank-two string-Carroll limit of the canonical stress--spin Ward system for probe matter on a fixed torsion-free background. Beyond rank-one Carroll spin hydrodynamics, this retains the coupled Riemann--spin force, a second longitudinal momentum projection, and the longitudinal boost bivector allowed only by a two-dimensional longitudinal kernel. We first extract the term linear in an independent spin amplitude $\varsigma$ and then expand in the near-horizon parameter $\lambda=\epsilon^2$, avoiding contamination by the omitted $\mathcal{O}(\varsigma^2)$ thermodynamic spin feedback. For finite mixed stress and $S^{\lambda\mu\nu}=\mathcal{O}(\varsigma\epsilon^p)$, $p=0$ is the unique sector in which the induced stress and curvature source balance at the first generic Ward order. On the regular branch of a smooth nonextremal static spherical outer horizon, locally normalized longitudinal and transverse spin amplitudes obey the same dilution law, while the curvature force first appears at $\mathcal{O}(\varsigma\lambda^2)$. For a general barotrope, the stationary system reduces to one thermodynamic quadrature and one local inverse. For the affine constant-sound-speed family $p=\alpha\mathcal E+\Pi$, $0\leq\alpha<1$, we obtain an explicit closed-form parametric solution, leading profiles for both spin channels, and fixed-radius, linear-spin corrections to enthalpy, pressure, and rapidity away from the sonic point. The response is governed by $\mathcal R^{(L)}_h=(f_2+3f_1\psi_1)/2$ and the longitudinal spin flux. We prove that the stationary closed-sphere Killing flux is invariant under regular pseudo-gauge improvements. In Reissner--Nordstr\"om, the response vanishes at the radial-tidal inversion $|Q|/M=2\sqrt{2}/3$; in the Einstein--Maxwell--dilaton family, its zero is $R_-/R_+=(1+a^2)/2$ for $0\leq a<1$. The Frenkel limit removes this response and the longitudinal boost sector.
\end{abstract}


\maketitle

\section{Introduction}
\label{sec:introduction}

Carrollian geometry arises in ultra-relativistic limits and on null hypersurfaces \cite{Henneaux:1979vn,Duval:2014uoa,Bagchi:2025vri}. Black-hole horizons have an intrinsic Carrollian geometry, and the null Raychaudhuri and Damour equations take the form of Carrollian conservation laws \cite{Donnay:2019jiz}. Projected Einstein equations on a stretched horizon also give Carrollian-fluid conservation laws \cite{Freidel:2022vjq}. Anyonic point-particle models have been used to study spin-dependent motion on Carrollian black-hole horizons \cite{Marsot:2022imf,Gray:2022svz}. These models describe horizon-confined particle kinematics rather than a hydrodynamic stress--spin Ward system in the two-longitudinal near-horizon geometry.

Near a nonextremal horizon, the rank-two string-Carroll expansion keeps two longitudinal directions instead of one. For Schwarzschild, it splits the metric and string sigma model into a two-dimensional Rindler sector and a transverse sphere \cite{Bagchi:2023cfp}. The same method applies to Schwarzschild, Reissner--Nordstr\"om, and Kerr outer horizons \cite{Bagchi:2024rje}. Reference~\cite{Bagchi:2026bnh} treats generic nonextremal black objects and particle and scalar probes on fixed backgrounds. Subleading orders in Carroll expansions can contain fluid variables absent at leading order \cite{Hansen:2021fxi,Kolekar:2024cfg}. Extremal horizons need a different scaling \cite{Kunduri:2013gce,Shinde:2026ern}.

Relativistic spin hydrodynamics uses coupled stress and spin currents. Their Ward identities follow by varying independent vielbein and spin-connection sources \cite{Hehl:1976kj,Hongo:2021ona,Gallegos:2022jow}. Pseudo-gauge transformations change both local currents \cite{Speranza:2020ilk}, and on a curved background the canonical stress identity contains a Riemann--spin term \cite{Chiarini:2024cuv}. A linear spin equation of state with one susceptibility does not satisfy the distinct stability signs of its electric- and magnetic-like sectors; a generalized Frenkel relation instead assigns them independent susceptibilities \cite{Daher:2024sce}. Perfect-spin-fluid thermodynamics also shows that spin-polarization feedback enters the stress and charge currents at quadratic order in the spin potential \cite{Florkowski:2024gtr,Drogosz:2024sff}, while pseudo-gauge choices affect the thermodynamic identification of ideal spin variables \cite{Armas:2026tif}. Fully causal and symmetric-hyperbolic nondissipative formulations impose further conditions \cite{Abboud:2025psf}. We therefore keep independent boost and rotation susceptibilities only as a linear-spin algebraic sector, fix the canonical pseudo-gauge before contraction, and order the spin and near-horizon expansions independently.

Rank-one Carroll fluids follow from symmetry and thermodynamics \cite{Freidel:2022bai,Armas:2023dcz}. Shukla et al.\ add an ideal spin current, begin from separately conserved stress and spin currents, and apply the resulting rank-one equations to Bjorken and Gubser flows \cite{Shukla:2026chs}. Our spin-current convention differs from Ref.~\cite{Shukla:2026chs} by an overall sign, $S_{\rm here}^{\lambda\mu\nu}=-S_{\rm Shukla}^{\lambda\mu\nu}$, and therefore $\Omega_{\rm here}^{\mu\nu}=-\Omega_{\rm Shukla}^{\mu\nu}$ for the convective ansatz. The sign difference is invisible in their symmetric ideal sector, where the stress and spin currents are separately conserved, but it matters when we compare the canonical curvature force. The present formulation extends that setup in four respects relevant here. We contract the coupled canonical Ward identities rather than separate conservation laws. We retain the curved-background Riemann--spin source. The rank-two degenerate longitudinal subspace supplies a second longitudinal momentum equation and one longitudinal boost bivector. The stationary spherical sector admits the analytic curvature response derived below. These statements concern the Ward system and the dimension of the degenerate longitudinal subspace. We do not claim that every constitutive model or solution of Ref.~\cite{Shukla:2026chs} is contained as a special case.

String-Carroll black-hole studies derive the rank-two near-horizon geometry and analyze particle or scalar probes \cite{Bagchi:2023cfp,Bagchi:2024rje,Bhattacharya:2026ueg,Bagchi:2026bnh}. Carroll-fluid theories usually use one degenerate longitudinal direction \cite{Freidel:2022bai,Armas:2023dcz,Shukla:2026chs}. Spin theories on curved backgrounds retain the geometric spin coupling but do not take the rank-two near-horizon limit \cite{Hongo:2021ona,Gallegos:2022jow,Chiarini:2024cuv}. Reference~\cite{Shukla:2026chs} contracts separately conserved ideal stress and spin currents after setting physical torsion to zero; it does not contract a canonical stress identity with its curvature source. The specific advance pursued here is therefore not a new horizon geometry or a new curved-space Ward identity. It is the contraction of that canonical Ward system in a two-longitudinal horizon geometry, together with the response carried by the longitudinal spin bivector.

We combine the curvature-sourced canonical stress identity with the rank-two degenerate longitudinal subspace and an unconstrained spin sector in one near-horizon contraction. This two-dimensional subspace gives the second longitudinal momentum equation, while the antisymmetric spin density supplies the longitudinal boost channel. We introduce an independent dimensionless spin amplitude $\varsigma$ and write $S^{\lambda\mu\nu}=\mathcal{O}(\varsigma\epsilon^p)$. With $T^\mu{}_{\nu}=\mathcal{O}(1)$, $p=0$ is the balanced interacting sector, $p>0$ is passive at the first singular order, and $p<0$ requires additional singular matter terms. This is a first-generic-order classification: a fractional $p$ entails a nonanalytic contraction, and accidental zeros can postpone the first nonzero equation. We extract the coefficient linear in $\varsigma$ before expanding in $\lambda$. On the regular spherical branch, the negative-power terms vanish and the first nonzero spin-dependent force occurs at $\mathcal{O}(\varsigma\lambda^2)$. Ingoing coordinates select this branch for a smooth static spherical outer horizon in the areal-radius gauge used below.

For a general stationary barotrope, the leading equations reduce analytically to a thermodynamic quadrature and the inverse of one monotone equation-of-state function. The affine family $p=\alpha\mathcal E+\Pi$ has a fully explicit parametric solution up to its sonic point and contains the usual linear barotrope at $\Pi=0$. The exact convective spin equation on the unexpanded spherical geometry conserves the ratio of locally normalized channel amplitudes along every radial trajectory. The ratio in a fixed horizon-normalized bivector basis changes by $(N_\epsilon/N_h)(r_h/r_\epsilon)^2$; this is a basis conversion, not an independent transport coefficient. At $\mathcal{O}(\varsigma\lambda^2)$, the longitudinal curvature scalar drives a matched correction to the matter Killing-energy flux. We integrate that equation and give both the response-dressed parametric solution and explicit linear-spin corrections at fixed radius. The combined stress--spin Killing current supplies a conserved closed-sphere flux whose pseudo-gauge invariance is proved below. In Reissner--Nordstr\"om, the response coefficient vanishes at $|Q|/M=2\sqrt{2}/3$. This charge is the known horizon value at which the radial tidal eigenvalue changes sign \cite{Crispino:2016rnm}; its appearance in spin-induced flow response is the result relevant here. In the Einstein--Maxwell--dilaton family \cite{Gibbons:1987ps,Garfinkle:1990qj}, the zero depends on the coupling $a$ and exists only for $0\leq a<1$.

We treat the spinful fluid as probe matter on a fixed torsion-free background. The physical connection is Levi--Civita and the local equations are written in the canonical pseudo-gauge. The electric/boost and magnetic/rotation susceptibilities obey necessary low-momentum signs, but the algebraic truncation is not a complete causal initial-value theory. Our result is complete only as the coefficient linear in $\varsigma$ within an ideal stress closure with no explicit $\mathcal{O}(\varsigma)$ spin-potential term. Quadratic spin thermodynamics, derivative transport, and the sonic layer are outside that statement. The spherical application assumes a smooth areal-radius coordinate and the outer-horizon orientation $F'(r_h)>0$. We exclude inner and cosmological horizons, rotation, extremality, physical torsion, dissipation, backreaction, and global accretion.

Sections~\ref{sec:curved-spin-rank-two} and~\ref{sec:string-carroll-spin} derive the Ward identities and their rank-two limit. Sections~\ref{sec:sss-horizon} and~\ref{sec:analytic-spin-transport} apply them to spherical horizons, solve the leading stationary system, and integrate its matched $\mathcal{O}(\varsigma\lambda^2)$ response. Section~\ref{sec:black-hole-applications} treats Schwarzschild, Reissner--Nordstr\"om, and Einstein--Maxwell--dilaton black holes. The appendices give algebraic checks.

\section{Curved spin Ward identities and rank-two geometry}
\label{sec:curved-spin-rank-two}

We use Lorentzian signature $(-,+,\ldots,+)$ and natural units $c=\hbar=k_{\rm B}=1$. Greek indices $\mu,\nu,\ldots=0,\ldots,d-1$ label spacetime components, and capital Latin indices $I,J,\ldots$ label tangent-space components. Antisymmetrization and symmetrization have unit weight. We define $A^{[\mu\nu]}=(A^{\mu\nu}-A^{\nu\mu})/2$ and $A^{(\mu\nu)}=(A^{\mu\nu}+A^{\nu\mu})/2$. Our curvature convention is
\begin{equation}
  R^{\rho}{}_{\sigma\mu\nu}=2\partial_{[\mu}\Gamma^{\rho}_{\nu]\sigma}+2\Gamma^{\rho}_{[\mu|\lambda|}\Gamma^{\lambda}_{\nu]\sigma}, \label{eq:riemann-convention}
\end{equation}
so the Levi--Civita derivative satisfies $[\nabla_\mu,\nabla_\nu]V^\rho=R^\rho{}_{\sigma\mu\nu}V^\sigma$.

Let $\widehat E_\mu{}^I$ be the physical vielbein, $\widehat E^\mu{}_I$ its inverse, and $\eta_{IJ}=\mathrm{diag}(-1,+1,\ldots,+1)$. They obey $g_{\mu\nu}=\widehat E_\mu{}^I\widehat E_\nu{}^J\eta_{IJ}$, $\widehat E^\mu{}_I\widehat E_\mu{}^J=\delta_I{}^J$, and $\widehat E^\mu{}_I\widehat E_\nu{}^I=\delta^\mu{}_\nu$. Before setting torsion to zero, we treat $\widehat E_\mu{}^I$ and the metric-compatible spin connection $\omega_\mu{}^{IJ}=-\omega_\mu{}^{JI}$ as independent sources. The tetrad postulate is
\begin{equation}
  \mathcal D_\mu\widehat E_\nu{}^I=\partial_\mu\widehat E_\nu{}^I+\omega_\mu{}^I{}_J\widehat E_\nu{}^J-\Gamma^\rho{}_{\mu\nu}\widehat E_\rho{}^I=0,
\end{equation}
where $\mathcal D_\mu V^I=\partial_\mu V^I+\omega_\mu{}^I{}_J V^J$. The torsion and Lorentz curvature are

\begin{align}
  \mathcal T^I{}_{\mu\nu}&=2\partial_{[\mu}\widehat E_{\nu]}{}^I+2\omega_{[\mu}{}^I{}_J\widehat E_{\nu]}{}^J,\\ \mathcal R^{IJ}{}_{\mu\nu}&=2\partial_{[\mu}\omega_{\nu]}{}^{IJ}+2\omega_{[\mu}{}^I{}_K\omega_{\nu]}{}^{KJ},
\end{align}
and $\mathcal G_\mu\equiv\mathcal T^\nu{}_{\nu\mu}$ is the torsion trace.

For a matter generating functional $W[\widehat E,\omega]$, we define the canonical pseudo-gauge through
\begin{equation}
  \delta W=\int d^d x\,e\left(T^\mu{}_I\,\delta\widehat E_\mu{}^I-\frac12 S^\mu{}_{IJ}\,\delta\omega_\mu{}^{IJ}\right), \label{eq:source-currents}
\end{equation}
where $e=|\det\widehat E_\mu{}^I|$. The Palatini formulation treats the vielbein and spin connection as independent sources \cite{Hehl:1976kj,Hongo:2021ona,Gallegos:2022jow}. The spin current is antisymmetric only in $I,J$ and has no corresponding symmetry in its flux index. We do not apply a Belinfante--Rosenfeld improvement. We hold the pseudo-gauge fixed because pseudo-gauge transformations on curved spacetime also change curvature terms \cite{Speranza:2020ilk,Chiarini:2024cuv}.

Under a local Lorentz transformation $\beta_{\rm L}^{IJ}=-\beta_{\rm L}^{JI}$, we take $\delta_{\rm L}\widehat E_\mu{}^I=-\beta_{\rm L}^I{}_J\widehat E_\mu{}^J$ and $\delta_{\rm L}\omega_\mu{}^{IJ}=\mathcal D_\mu\beta_{\rm L}^{IJ}$. Integration by parts on a Riemann--Cartan background gives
\begin{equation}
  \int d^d x\,e\,V^\mu\mathcal D_\mu\phi=-\int d^d x\,e\,(\mathcal D_\mu-\mathcal G_\mu)V^\mu\,\phi,
\end{equation}
up to a boundary term. For a diffeomorphism generated by $\xi^\mu$, a compensating local Lorentz transformation gives

\begin{align}
  \delta_\xi\widehat E_\mu{}^I&=\mathcal D_\mu\xi^I+\xi^\nu\mathcal T^I{}_{\nu\mu},\\ \delta_\xi\omega_\mu{}^{IJ}&=\xi^\nu\mathcal R^{IJ}{}_{\nu\mu}.
\end{align}
Here $\xi^I=\xi^\mu\widehat E_\mu{}^I$. Substitution into Eq.~\eqref{eq:source-currents} yields

\begin{align}
  \delta_{\rm L} W&=\int d^d x\,e\,\beta_{\rm L}^{IJ}\left[T_{[IJ]}+\frac12(\mathcal D_\mu-\mathcal G_\mu)S^\mu{}_{IJ}\right],\\ \delta_\xi W&=\int d^d x\,e\,\xi^I\left[-(\mathcal D_\mu-\mathcal G_\mu)T^\mu{}_I-T^\mu{}_J\mathcal T^J{}_{\mu I}\right.\\ &\hspace{3.4cm}\left.+\frac12 S^\mu{}_{JK}\mathcal R^{JK}{}_{\mu I}\right]. \nonumber
\end{align}
Because $\beta_{\rm L}^{IJ}$ and $\xi^I$ are arbitrary, local Lorentz and diffeomorphism invariance give the Riemann--Cartan Ward identities
\begin{subequations}\label{eq:rc-spin-ward}
  \begin{align}
    (\mathcal D_\mu-\mathcal G_\mu)T^\mu{}_I&=-T^\mu{}_J\mathcal T^J{}_{\mu I}+\frac12 S^\mu{}_{JK}\mathcal R^{JK}{}_{\mu I},\\ (\mathcal D_\mu-\mathcal G_\mu)S^\mu{}_{IJ}&=-2T_{[IJ]}.
  \end{align}
\end{subequations}
Here $\mathcal T^I{}_{\mu J}\equiv\mathcal T^I{}_{\mu\nu}\widehat E^\nu{}_J$, $\mathcal R^{IJ}{}_{\mu K}\equiv\mathcal R^{IJ}{}_{\mu\nu}\widehat E^\nu{}_K$, and $T_{IJ}\equiv\widehat E_{\mu I}T^\mu{}_J$. Equation~\eqref{eq:rc-spin-ward} describes neutral matter, so it contains no external gauge force. Its source structure agrees with Ward identities derived from independent vielbein and spin-connection variations \cite{Hongo:2021ona,Gallegos:2022jow}.

Set the background torsion to zero and define the spacetime currents by
\begin{equation}
  \begin{gathered}
    T^{\mu\nu}=T^\mu{}_I\widehat E^{\nu I},\\
    S^{\lambda\mu\nu}=S^\lambda{}_{IJ}\widehat E^{\mu I}\widehat E^{\nu J}.
  \end{gathered}
\end{equation}
Using the pair symmetry of the Levi--Civita Riemann tensor, the curvature term becomes
\begin{align}
  \frac12 S^\mu{}_{IJ}\mathcal R^{IJ}{}_{\mu K}\widehat E^{\nu K}&=\frac12 S^{\alpha\beta\gamma}R_{\beta\gamma\alpha}{}^\nu\nonumber\\ &=-\frac12 R^\nu{}_{\alpha\beta\gamma}S^{\alpha\beta\gamma}.
\end{align}

The torsion-free Ward identities are
\begin{equation}
  \begin{aligned}\nabla_\mu T^{\mu\nu}&=-\frac12 R^\nu{}_{\alpha\beta\gamma}S^{\alpha\beta\gamma},\\ \nabla_\lambda S^{\lambda\mu\nu}&=-2T^{[\mu\nu]}.
  \end{aligned} \label{eq:lc-spin-ward}
\end{equation}
For the convention in Eq.~\eqref{eq:riemann-convention}, these signs match the covariant angular-momentum derivation in Ref.~\cite{Chiarini:2024cuv}.

The two Ward identities also determine the conserved current associated with a Killing vector $\xi^\mu$. Define
\begin{equation}
  {\cal I}_\xi^\mu
  \equiv T^\mu{}_{\nu}\xi^\nu
  +\frac12S^{\mu\alpha\beta}\nabla_\alpha\xi_\beta.
  \label{eq:killing-spin-current}
\end{equation}
Using $\nabla_{(\mu}\xi_{\nu)}=0$, $\nabla_\mu\nabla_\alpha\xi_\beta=R_{\beta\alpha\mu\gamma}\xi^\gamma$, and Eq.~\eqref{eq:lc-spin-ward}, one obtains
\begin{equation}
  \nabla_\mu{\cal I}_\xi^\mu=0.
\end{equation}
The antisymmetric-stress term from the first contribution cancels the divergence of the spin term, while the two curvature terms cancel by the Riemann symmetries. The closed-surface statement can be proved directly. Let a pseudo-gauge improvement be generated by $\Phi^{\lambda\mu\nu}=-\Phi^{\lambda\nu\mu}$ and define
\begin{equation}
  \begin{gathered}
    B^{\lambda\mu\nu}\equiv\frac12\left(
    \Phi^{\lambda\mu\nu}+\Phi^{\mu\nu\lambda}
  +\Phi^{\nu\mu\lambda}\right),\\
    B^{\lambda\mu\nu}=-B^{\mu\lambda\nu}.
  \end{gathered}
\end{equation}
For $T'^{\mu\nu}=T^{\mu\nu}+\nabla_\lambda B^{\lambda\mu\nu}$ and $S'^{\lambda\mu\nu}=S^{\lambda\mu\nu}-\Phi^{\lambda\mu\nu}$, the Killing equation gives
\begin{equation}
  \begin{gathered}
    {\cal I}_\xi^{\prime\mu}-{\cal I}_\xi^\mu
  =\nabla_\lambda U_\xi^{\lambda\mu},\\
    U_\xi^{\lambda\mu}\equiv B^{\lambda\mu}{}_{\nu}\xi^\nu
  =-U_\xi^{\mu\lambda}.
  \label{eq:pseudogauge-killing-superpotential}
  \end{gathered}
\end{equation}
Indeed, the contraction $B^{\lambda\mu\nu}\nabla_\lambda\xi_\nu=-\Phi^{\mu\lambda\nu}\nabla_\lambda\xi_\nu/2$ cancels the transformed spin term. For a stationary radial flux, \begin{equation}
  \sqrt{-g}\,\nabla_\lambda U_\xi^{\lambda\rho}
  =\partial_\lambda\!\left(\sqrt{-g}\,U_\xi^{\lambda\rho}\right).
\end{equation}
The $v$ derivative vanishes, $U_\xi^{\rho\rho}=0$, and the angular derivatives integrate to zero on a closed two-sphere. Thus the flux of ${\cal I}_\xi^\mu$ is invariant under stationary improvements for which $U_\xi^{\lambda\mu}$ is regular and single-valued on the sphere \cite{Speranza:2020ilk,Chiarini:2024cuv}. These assumptions, rather than the separate local stress or spin term, define the pseudo-gauge-independent check used below.

For the rank-two limit, split the tangent index as $I=(A,a)$, where $A,B=0,1$ are longitudinal and $a,b=2,\ldots,d-1$ are transverse. We use $\eta_{AB}=\mathrm{diag}(-1,+1)$ and the Euclidean transverse metric $\delta_{ab}$. The two-dimensional longitudinal split is the string ($p=1$) case of $p$-brane Carroll geometry, whose defining geometric structures are complementary degenerate tensors \cite{Bergshoeff:2023ogz}. With a dimensionless parameter $\epsilon>0$, scale the frame as
\begin{equation}
  \begin{gathered}
    \widehat E_\mu{}^A=\epsilon E_\mu{}^A,\\
    \widehat E_\mu{}^a=E_\mu{}^a. \label{eq:rank-two-frame-scaling}
  \end{gathered}
\end{equation}

The inverse frame then satisfies $\widehat E^\mu{}_A=\epsilon^{-1}E^\mu{}_A$ and $\widehat E^\mu{}_a=E^\mu{}_a$. The metric and inverse metric take the form
\begin{subequations}\label{eq:rank-two-metric}
  \begin{align}
  &\begin{gathered}
      g_{\mu\nu}(\epsilon)=\epsilon^2 V_{\mu\nu}+\Pi_{\mu\nu},\\
      g^{\mu\nu}(\epsilon)=\epsilon^{-2}V^{\mu\nu}+\Pi^{\mu\nu},
    \end{gathered}\\
  &\begin{gathered}
      V_{\mu\nu}=\eta_{AB}E_\mu{}^A E_\nu{}^B,\\
      V^{\mu\nu}=\eta^{AB}E^\mu{}_A E^\nu{}_B,
    \end{gathered}\\
  &\begin{gathered}
      \Pi_{\mu\nu}=\delta_{ab}E_\mu{}^a E_\nu{}^b,\\
      \Pi^{\mu\nu}=\delta^{ab}E^\mu{}_a E^\nu{}_b.
    \end{gathered}
\end{align}
\end{subequations}

This scaling is used in two-longitudinal near-horizon string expansions \cite{Bagchi:2023cfp,Bagchi:2024rje}. The inverse-frame relations imply
\begin{subequations}\label{eq:rank-two-projectors}
  \begin{align}
    V_{\mu\rho}\Pi^{\rho\nu}&=0,\\V^{\mu\rho}\Pi_{\rho\nu}&=0,\\ P_\parallel{}^\mu{}_\nu&\equiv V^{\mu\rho}V_{\rho\nu}=E^\mu{}_A E_\nu{}^A,\\ P_\perp{}^\mu{}_\nu&\equiv\Pi^{\mu\rho}\Pi_{\rho\nu}=E^\mu{}_a E_\nu{}^a,\\ P_\parallel+P_\perp&=\mathbf 1,\\P_\parallel P_\perp&=0,\\ P_\parallel^2&=P_\parallel,\\P_\perp^2&=P_\perp.
  \end{align}
\end{subequations}
The two projectors have traces $2$ and $d-2$.

Assume an even analytic frame expansion,
\begin{equation}
  E_\mu{}^I=e_\mu{}^I+\epsilon^2e_{(2)\mu}{}^I+\mathcal{O}(\epsilon^4).
\end{equation}
The leading frame tensors are
\begin{subequations}\label{eq:leading-frame-complements}
  \begin{align}
  &\begin{gathered}
      v_{\mu\nu}\equiv\eta_{AB}e_\mu{}^A e_\nu{}^B,\\
      v^{\mu\nu}\equiv\eta^{AB}e^\mu{}_A e^\nu{}_B,
    \end{gathered}\\
  &\begin{gathered}
      h_{\mu\nu}\equiv\delta_{ab}e_\mu{}^a e_\nu{}^b,\\
      h^{\mu\nu}\equiv\delta^{ab}e^\mu{}_a e^\nu{}_b.
    \end{gathered}
\end{align}
\end{subequations}
Here $v_{\mu\nu}$ and $v^{\mu\nu}$ are the covariant and contravariant longitudinal tensors, while $h_{\mu\nu}$ and $h^{\mu\nu}$ are their transverse counterparts. Each tensor is degenerate on the complementary subspace. At each point $p$ of spacetime, define the null space of the covariant transverse metric by
\begin{equation}
  \ker h_p\equiv
  \left\{X\in T_p\mathcal M\mid h_{\mu\nu}X^\nu=0\right\}.
\end{equation}
Equivalently, $h_p(X,Y)=0$ for every $Y\in T_p\mathcal M$. Since $h_{\mu\nu}e^\nu{}_A=0$ for $A=0,1$, this null space is the longitudinal two-plane spanned by the vectors $e_A=e^\mu{}_A\partial_\mu$. Thus $\dim\ker h_p=2$. Physically, $h_{\mu\nu}$ measures only transverse distances and is insensitive to displacements along this longitudinal plane. The tensor $v_{\mu\nu}$ supplies the Lorentzian geometry on that plane. In the spherical near-horizon application, the plane is the ingoing time--radial sector. The tensor $h^{\mu\nu}$ is frame-defined and is not an ordinary inverse of the rank-$(d-2)$ tensor $h_{\mu\nu}$. They satisfy
\begin{subequations}\label{eq:leading-frame-projectors}
  \begin{align}
  &\begin{gathered}
      v_{\mu\rho}h^{\rho\nu}=0,\\
      v^{\mu\rho}h_{\rho\nu}=0,
    \end{gathered}\\
  &\begin{gathered}
      p_\parallel{}^\mu{}_\nu=v^{\mu\rho}v_{\rho\nu},\\
      p_\perp{}^\mu{}_\nu=h^{\mu\rho}h_{\rho\nu},
    \end{gathered}\\
p_\parallel{}^\mu{}_\nu+p_\perp{}^\mu{}_\nu&=\delta^\mu{}_\nu.
\end{align}
\end{subequations}
The tensors $p_\parallel{}^\mu{}_\nu$ and $p_\perp{}^\mu{}_\nu$ are the leading longitudinal and transverse projectors, respectively. The notation $h_{\mu\nu}$, $h^{\mu\nu}$, and $v^{\mu\nu}$ follows standard string-Carroll usage \cite{Bagchi:2024rje}. We use $v_{\mu\nu}$ for the covariant longitudinal complement rather than the common $\tau_{\mu\nu}$ because $\tau^\mu{}_\nu$ below denotes the leading mixed stress tensor. The inverse-frame expansion gives $V_{\mu\nu}=v_{\mu\nu}+\mathcal{O}(\epsilon^2)$, $V^{\mu\nu}=v^{\mu\nu}+\mathcal{O}(\epsilon^2)$, $\Pi_{\mu\nu}=h_{\mu\nu}+\epsilon^2\pi_{(2)\mu\nu}+\mathcal{O}(\epsilon^4)$, and $\Pi^{\mu\nu}=h^{\mu\nu}+\mathcal{O}(\epsilon^2)$. The metric is then
\begin{align}
  g_{\mu\nu}&=h_{\mu\nu}+\epsilon^2\left(v_{\mu\nu}+\pi_{(2)\mu\nu}\right)+\mathcal{O}(\epsilon^4)\nonumber\\ &\equiv g_{(0)\mu\nu}+\epsilon^2g_{(2)\mu\nu}+\mathcal{O}(\epsilon^4).
\end{align}
The leading covariant metric $h_{\mu\nu}$ has constant rank $d-2$, so its null spaces form a smooth two-dimensional longitudinal subbundle. The inverse of the full metric is a Laurent series, $g^{\mu\nu}=\epsilon^{-2}v^{\mu\nu}+\mathcal{O}(\epsilon^0)$, rather than a Taylor series. Such even expansions are standard in Carroll limits \cite{Hansen:2021fxi}.

For finite $\epsilon$, the full metric is Lorentzian. Scale the mixed Lorentz parameter as $\beta_{\rm L}^A{}_a=-\epsilon\Lambda^A{}_a$. As $\epsilon\to0$, the mixed Lorentz transformation becomes the local string-Carroll boost \cite{Bagchi:2023cfp,Bagchi:2024rje},
\begin{subequations}\label{eq:string-carroll-boost}
  \begin{align}
  &\begin{gathered}
      \delta_{\rm B}e_\mu{}^A=\Lambda^A{}_a e_\mu{}^a,\\
      \delta_{\rm B}e_\mu{}^a=0,
    \end{gathered}\\
  &\begin{gathered}
      \delta_{\rm B}e^\mu{}_A=0,\\
      \delta_{\rm B}e^\mu{}_a=-\Lambda_a{}^A e^\mu{}_A.
    \end{gathered}
\end{align}
\end{subequations}
Equation~\eqref{eq:string-carroll-boost} does not include the independent longitudinal $S\mathcal{O}(1,1)$ transformation or transverse $S\mathcal{O}(d-2)$ rotation. Indices on $\Lambda^A{}_a$ are moved with $\eta_{AB}$ and $\delta_{ab}$. The tensors $v^{\mu\nu}$ and $h_{\mu\nu}$ are boost invariant, while their complements transform as

\begin{align}
  \delta_{\rm B}v_{\mu\nu}&=2\eta_{AB}\Lambda^A{}_a e^B{}_{(\mu}e^a{}_{\nu)},\\ \delta_{\rm B}h^{\mu\nu}&=-2\delta^{ab}\Lambda^A{}_a e_A{}^{(\mu}e_b{}^{\nu)}.
\end{align}
These transformations preserve the leading projector relations in Eq.~\eqref{eq:leading-frame-projectors}.

Keep the flux index of the spin current separate from its antisymmetric pair. Using the leading coframe, define $T^{IJ}=e_\mu{}^I e_\nu{}^J T^{\mu\nu}$ and

\begin{align}
  S^{\lambda|AB}&=e_\mu{}^A e_\nu{}^B S^{\lambda\mu\nu},\\ S^{\lambda|Aa}&=e_\mu{}^A e_\nu{}^a S^{\lambda\mu\nu},\\ S^{\lambda|ab}&=e_\mu{}^a e_\nu{}^b S^{\lambda\mu\nu}.
\end{align}

The inverse relation is
\begin{equation}
  S^{\lambda\mu\nu}=e_A{}^\mu e_B{}^\nu S^{\lambda|AB}+2e_A{}^{[\mu}e_a{}^{\nu]}S^{\lambda|Aa}+e_a{}^\mu e_b{}^\nu S^{\lambda|ab}.
\end{equation}

For $n_\perp=d-2$ transverse directions, the antisymmetric pair contains

\begin{align}
\binom{2}{2}+2n_\perp+\binom{n_\perp}{2}&=\binom{d}{2},\\
  &\begin{gathered}
      d=4,\\
      6=1_{\parallel}+4_{\rm mix}+1_{\perp}.
    \end{gathered}
\end{align}
The $1+4+1$ count applies only to the antisymmetric pair. Before constitutive or spacetime-symmetry restrictions, a general four-dimensional current $S^{\lambda\mu\nu}$ has $4\times6=24$ components.

In the unscaled tangent basis, $\widetilde\eta_{IJ}=\mathrm{diag}(\epsilon^2\eta_{AB},\delta_{ab})$. Lowering one longitudinal tangent index adds a factor $\epsilon^2$. This index weight does not determine the amplitude of $T^{IJ}$ or $S^{\lambda|IJ}$.

The boost mixes the projected currents in a triangular form.
\begin{subequations}\label{eq:spin-block-boost}
  \begin{align}
    \delta_{\rm B}S^{\lambda|AB}&=\Lambda^A{}_a S^{\lambda|aB}+\Lambda^B{}_a S^{\lambda|Aa},\\
    \delta_{\rm B}S^{\lambda|Aa}&=\Lambda^A{}_b S^{\lambda|ba},\\
    \delta_{\rm B}S^{\lambda|ab}&=0.
  \end{align}
\end{subequations}
and
\begin{subequations}\label{eq:stress-block-boost}
  \begin{align}
    \delta_{\rm B}T^{AB}&=\Lambda^A{}_aT^{aB}+\Lambda^B{}_aT^{Aa},\\
    \delta_{\rm B}T^{Aa}&=\Lambda^A{}_bT^{ba},\\
    \delta_{\rm B}T^{aB}&=\Lambda^B{}_bT^{ab},\\
    \delta_{\rm B}T^{ab}&=0.
  \end{align}
\end{subequations}
The flux index $\lambda$ remains a spacetime index in these relations.

If $S^{\lambda\mu\nu}=0$, the equations give a symmetric conserved stress tensor. On a flat background, the curvature force vanishes, but spin can still exchange with the antisymmetric canonical stress. For a rank-one split, $v^{\mu\nu}=-k^\mu k^\nu$, so the longitudinal--longitudinal spin component is absent. This kinematics matches rank-one Carroll-fluid theories \cite{Freidel:2022bai,Armas:2023dcz,Shukla:2026chs}. A dynamical comparison also requires the same connection, pseudo-gauge, Ward identities, and constitutive assumptions.

\section{String-Carroll spin transport}
\label{sec:string-carroll-spin}

Insert the rank-two metric in Eq.~\eqref{eq:rank-two-metric} into the torsion-free Ward identities in Eq.~\eqref{eq:lc-spin-ward}. We first lower the free index of the stress equation,
\begin{equation}
  \nabla_\mu T^\mu{}_{\nu} =-\frac12 R_{\nu\alpha\beta\gamma} S^{\alpha\beta\gamma}. \label{eq:sc-mixed-ward}
\end{equation}
The curvature term introduces no additional power of $\epsilon$.

Define the derivatives of the two metric blocks by

\begin{align}
  V_{\lambda\mu\nu} &=\frac12\left(\partial_\mu V_{\lambda\nu} +\partial_\nu V_{\lambda\mu} -\partial_\lambda V_{\mu\nu}\right),\\ \Pi_{\lambda\mu\nu} &=\frac12\left(\partial_\mu \Pi_{\lambda\nu} +\partial_\nu \Pi_{\lambda\mu} -\partial_\lambda \Pi_{\mu\nu}\right).
\end{align}
Substituting $g_{\mu\nu}=\epsilon^2V_{\mu\nu}+\Pi_{\mu\nu}$ and $g^{\mu\nu}=\epsilon^{-2}V^{\mu\nu}+\Pi^{\mu\nu}$ into the Christoffel formula gives
\begin{subequations}\label{eq:sc-connection-exact}
  \begin{align}
    \Gamma^\rho{}_{\mu\nu} &=\epsilon^{-2}{\cal A}^\rho{}_{\mu\nu} +{\cal C}^\rho{}_{\mu\nu} +\epsilon^2{\cal B}^\rho{}_{\mu\nu},\\ {\cal A}^\rho{}_{\mu\nu} &=V^{\rho\lambda}\Pi_{\lambda\mu\nu},\\ {\cal C}^\rho{}_{\mu\nu} &=V^{\rho\lambda}V_{\lambda\mu\nu} +\Pi^{\rho\lambda}\Pi_{\lambda\mu\nu},\\ {\cal B}^\rho{}_{\mu\nu} &=\Pi^{\rho\lambda}V_{\lambda\mu\nu}.
  \end{align}
\end{subequations}

This algebraic split is the connection decomposition of the two-longitudinal string-Carroll metric \cite{Bagchi:2023cfp,Bagchi:2024rje}. The inverse-frame identities also give $\mathcal A^\rho{}_{\rho\nu}=0$.

The frame fields are analytic in $\epsilon^2$, so $\mathcal A$, $\mathcal C$, and $\mathcal B$ have even expansions. Write $\mathcal A=\sum_{j\geq0}\epsilon^{2j}{\cal A}_{(2j)}$ and use the same notation for $\mathcal C$ and $\mathcal B$. Then
\begin{equation}
  \Gamma^\rho{}_{\mu\nu} =\epsilon^{-2}\Gamma^{(-2)\rho}{}_{\mu\nu} +\Gamma^{(0)\rho}{}_{\mu\nu} +\epsilon^2\Gamma^{(2)\rho}{}_{\mu\nu} +\mathcal{O}(\epsilon^4), \label{eq:sc-connection-series}
\end{equation}
where
\begin{align}
  \Gamma^{(-2)}&={\cal A}_{(0)},\\ \Gamma^{(0)}&={\cal A}_{(2)}+{\cal C}_{(0)},\\ \Gamma^{(2)}&={\cal A}_{(4)}+{\cal C}_{(2)}+{\cal B}_{(0)}.
\end{align}

The trace satisfies $\Gamma^{(-2)\rho}{}_{\rho\nu}=0$. Under an $\epsilon$-independent coordinate transformation, the inhomogeneous connection term is $\mathcal{O}(\epsilon^0)$. Hence $\Gamma^{(-2)}$ transforms as a tensor, while $\Gamma^{(0)}$ transforms as a connection. For
\begin{equation}
  (X\star Y)^\rho{}_{\sigma\mu\nu} \equiv 2X^\rho{}_{[\mu|\lambda|}Y^\lambda{}_{\nu]\sigma},
\end{equation}
the Riemann convention in Eq.~\eqref{eq:riemann-convention} gives

\begin{align}
  R^{(-4)\rho}{}_{\sigma\mu\nu} &=(\Gamma^{(-2)}\star\Gamma^{(-2)}) ^\rho{}_{\sigma\mu\nu},\\ R^{(-2)\rho}{}_{\sigma\mu\nu} &=2\partial_{[\mu}\Gamma^{(-2)\rho}{}_{\nu]\sigma} +(\Gamma^{(-2)}\star\Gamma^{(0)}) ^\rho{}_{\sigma\mu\nu}\\ &\quad +(\Gamma^{(0)}\star\Gamma^{(-2)}) ^\rho{}_{\sigma\mu\nu}, \nonumber\\ R^{(0)\rho}{}_{\sigma\mu\nu} &=R^\rho{}_{\sigma\mu\nu}[\Gamma^{(0)}] +(\Gamma^{(-2)}\star\Gamma^{(2)}) ^\rho{}_{\sigma\mu\nu}\\ &\quad +(\Gamma^{(2)}\star\Gamma^{(-2)}) ^\rho{}_{\sigma\mu\nu}. \nonumber
\end{align}

The contravariant Riemann tensor can start at $\epsilon^{-4}$, but this term vanishes after lowering its first index. Each $\Gamma^{(-2)\rho}{}_{\mu\nu}$ has its upper index in the longitudinal image of $v^{\mu\nu}$, so $h_{\mu\rho}R^{(-4)\rho}{}_{\sigma\alpha\beta}=0$. Write the covariant metric as
\begin{equation}
  g_{\mu\nu}=h_{\mu\nu} +\epsilon^2 k_{\mu\nu} +\epsilon^4\ell_{\mu\nu} +\mathcal{O}(\epsilon^6), \label{eq:sc-covariant-metric-series}
\end{equation}
where $k_{\mu\nu}=g_{(2)\mu\nu}$ and $\ell_{\mu\nu}=g_{(4)\mu\nu}$. The lowered Riemann tensor is

\begin{align}
  R_{\nu\alpha\beta\gamma} &=\epsilon^{-2} {\mathfrak R}^{(-2)}_{\nu\alpha\beta\gamma} +{\mathfrak R}^{(0)}_{\nu\alpha\beta\gamma} +\mathcal{O}(\epsilon^2),\\ {\mathfrak R}^{(-2)}_{\nu\alpha\beta\gamma} &=h_{\nu\rho}R^{(-2)\rho}{}_{\alpha\beta\gamma} +k_{\nu\rho}R^{(-4)\rho}{}_{\alpha\beta\gamma},\\ {\mathfrak R}^{(0)}_{\nu\alpha\beta\gamma} &=h_{\nu\rho}R^{(0)\rho}{}_{\alpha\beta\gamma} +k_{\nu\rho}R^{(-2)\rho}{}_{\alpha\beta\gamma}\\ &\quad +\ell_{\nu\rho}R^{(-4)\rho}{}_{\alpha\beta\gamma}. \nonumber
\end{align}

The curvature force in Eq.~\eqref{eq:sc-mixed-ward} starts at $\epsilon^{-2}$ rather than $\epsilon^{-4}$. To separate small spin from the Carroll contraction, introduce a dimensionless spin amplitude $\varsigma$ and expand
\begin{equation}
\begin{split}
  T^\mu{}_{\nu}(\epsilon,\varsigma) &=\tau_{[0]}^\mu{}_{\nu}+\varsigma\tau_{[1]}^\mu{}_{\nu}+\epsilon^2\left(\tau_{(2)[0]}^\mu{}_{\nu}
  +\varsigma\tau_{(2)[1]}^\mu{}_{\nu}\right)\\
  &+\mathcal{O}(\epsilon^4)+\mathcal{O}(\varsigma^2),\\
  S^{\lambda\mu\nu}(\epsilon,\varsigma)
  &=\varsigma\epsilon^p\left(s^{\lambda\mu\nu}
    +\epsilon^2s_{(2)}^{\lambda\mu\nu}
  +\mathcal{O}(\epsilon^4)\right)\\
  &\quad+\mathcal{O}(\varsigma^2).
\end{split}
\end{equation}
Here $[0]$ and $[1]$ denote the spinless value and the derivative with respect to $\varsigma$ at $\varsigma=0$. The divergence of the induced stress $\tau_{[1]}$ starts at $\varsigma\epsilon^{-2}$ because of $\Gamma^{(-2)}$, while the curvature source is $\mathcal{O}(\varsigma\epsilon^{p-2})$ on a generic branch. The three scaling classes at the first generic Ward order are therefore
\begin{center}
  \footnotesize
  \renewcommand{\arraystretch}{1.18}
  \begin{tabular}{@{}c@{\hspace{0.7em}}c@{\hspace{0.7em}}c@{}}
    \hline
    $p$ & \parbox[c]{0.25\columnwidth}{\centering First curvature order} & \parbox[c]{0.48\columnwidth}{\centering Finite-stress interpretation}\\
    \hline
    $p>0$ & \parbox[c]{0.25\columnwidth}{\centering $\varsigma\epsilon^{p-2}$} & \parbox[c]{0.48\columnwidth}{\raggedright Spin is passive at $\epsilon^{-2}$.}\\
    $p=0$ & \parbox[c]{0.25\columnwidth}{\centering $\varsigma\epsilon^{-2}$} & \parbox[c]{0.48\columnwidth}{\raggedright Induced stress balances the source.}\\
    $p<0$ & \parbox[c]{0.25\columnwidth}{\centering $\varsigma\epsilon^{p-2}$} & \parbox[c]{0.48\columnwidth}{\raggedright A more singular stress sector is required.}\\
    \hline
  \end{tabular}
\end{center}
Thus $p=0$ is the unique interacting scaling subject to three stated requirements: $T^\mu{}_{\nu}=\mathcal{O}(1)$, a finite nonzero coefficient $S/\varsigma$, and a curvature term that participates in the first generic equation for $\tau_{[1]}$. We call it the balanced finite-current sector and set
\begin{equation}
  S^{\lambda\mu\nu}
  =\varsigma\left(s^{\lambda\mu\nu}
  +\epsilon^2s_{(2)}^{\lambda\mu\nu}+\mathcal{O}(\epsilon^4)\right)
  +\mathcal{O}(\varsigma^2).
  \label{eq:sc-spin-current-series}
\end{equation}
For the convective current used below, $\widehat u^\mu=\mathcal{O}(\epsilon^{-1})$ implies $\Omega^{\mu\nu}=\mathcal{O}(\epsilon^{p+1})$. The corresponding parent orthonormal components scale as
\begin{align}
  \Omega_{\rm phys}^{AB}&=\mathcal{O}(\epsilon^{p+3}),\\
  \Omega_{\rm phys}^{Aa}&=\mathcal{O}(\epsilon^{p+2}),\\
  \Omega_{\rm phys}^{ab}&=\mathcal{O}(\epsilon^{p+1}).
\end{align}
Thus every parent component vanishes for $p\geq0$, although its coefficient linear in $\varsigma$ can remain finite. This classification is a contraction statement, not a microscopic derivation of the spin density. It refers only to the first generic order: noninteger $p$ requires a fractional-power expansion, while vanishing geometric or constitutive coefficients can postpone the first nonzero term. The longitudinal, mixed, and transverse components of the $p=0$ sector form a closed set under local string-Carroll boosts. On the regular spherical branch used below, all negative-power coefficients vanish and the first spin-dependent force occurs at $\mathcal{O}(\varsigma\lambda^2)=\mathcal{O}(\varsigma\epsilon^4)$.

Define

\begin{align}
  {\cal L}_{-2}[X]_{\nu} &\equiv -\Gamma^{(-2)\rho}{}_{\mu\nu}X^\mu{}_{\rho},\\ {\cal Q}_{-2}[Y]^{\mu\nu} &\equiv \Gamma^{(-2)\mu}{}_{\lambda\rho}Y^{\lambda\rho\nu} +\Gamma^{(-2)\nu}{}_{\lambda\rho}Y^{\lambda\mu\rho}.
\end{align}
Write $g^{\mu\nu} =\epsilon^{-2}v^{\mu\nu}+\gamma^{\mu\nu}+\mathcal{O}(\epsilon^2)$ and let $\nabla^{(0)}$ be the covariant derivative built from $\Gamma^{(0)}$. Expanding both Ward identities gives
\begin{subequations}
  \begin{align}
    {\cal L}_{-2}[\tau_{[1]}]_{\nu}
    &=-\frac12 {\mathfrak R}^{(-2)}_{\nu\alpha\beta\gamma}
    s^{\alpha\beta\gamma},
    \label{eq:sc-ward-leading-stress}\\
    {\cal Q}_{-2}[s]^{\mu\nu}
    &=-2\tau_{[1]}^{[\mu}{}_{\rho}v^{\nu]\rho}.
    \label{eq:sc-ward-leading-spin}
  \end{align}
  \begin{align}
    \nabla^{(0)}_\mu\tau_{[1]}^\mu{}_{\nu}
    +{\cal L}_{-2}[\tau_{(2)[1]}]_{\nu}
    &=-\frac12 {\mathfrak R}^{(0)}_{\nu\alpha\beta\gamma}
    s^{\alpha\beta\gamma}\nonumber\\
    &\quad-\frac12 {\mathfrak R}^{(-2)}_{\nu\alpha\beta\gamma}
    s_{(2)}^{\alpha\beta\gamma},
    \label{eq:sc-ward-next-stress}\\
    \nabla^{(0)}_\lambda s^{\lambda\mu\nu}
    +{\cal Q}_{-2}[s_{(2)}]^{\mu\nu}
    &=-2\tau_{(2)[1]}^{[\mu}{}_{\rho}v^{\nu]\rho}
    -2\tau_{[1]}^{[\mu}{}_{\rho}\gamma^{\nu]\rho}.
    \label{eq:sc-ward-next-spin}
  \end{align}
\end{subequations}

Equations~\eqref{eq:sc-ward-leading-stress} and \eqref{eq:sc-ward-leading-spin} are the $\epsilon^{-2}$ equations for the coefficient linear in $\varsigma$, while Eqs.~\eqref{eq:sc-ward-next-stress} and \eqref{eq:sc-ward-next-spin} are its $\epsilon^0$ equations. The spinless equations are obtained by setting the right-hand sides to zero and replacing $[1]$ by $[0]$. This order of operations---first $\partial_\varsigma|_{0}$, then the $\epsilon$ expansion---is used throughout. We use an ideal long-wavelength closure and omit dissipative and derivative terms. Let $\widehat u^\mu$ be the unit timelike velocity, $g_{\mu\nu}\widehat u^\mu\widehat u^\nu=-1$, and define $U^\mu\equiv\epsilon\widehat u^\mu$. Its expansion is
\begin{subequations}\label{eq:sc-velocity-series}
  \begin{align}
    U^\mu &=u^\mu+\epsilon^2u_{(2)}^\mu+\mathcal{O}(\epsilon^4),\\ \epsilon^{-2}U_\mu &=\vartheta_\mu+\epsilon^2\vartheta_{(2)\mu} +\mathcal{O}(\epsilon^4).
  \end{align}
\end{subequations}
The velocity normalization and inverse metric give
\begin{equation}
  \begin{gathered}
    h_{\mu\nu}u^\nu=0,\\
    v^{\mu\nu}\vartheta_\nu=u^\mu,\\
    u^\mu\vartheta_\mu=-1. \label{eq:sc-velocity-constraints}
  \end{gathered}
\end{equation}

The covariant and contravariant velocity expansions are not independent. Using Eq.~\eqref{eq:sc-covariant-metric-series} in $U_\mu=g_{\mu\nu}U^\nu$ gives
\begin{equation}
  \vartheta_\mu=h_{\mu\nu}u_{(2)}^\nu+k_{\mu\nu}u^\nu.
  \label{eq:sc-leading-covariant-velocity}
\end{equation}
Conversely, expanding $U^\mu=g^{\mu\nu}U_\nu$ gives
\begin{equation}
  \begin{gathered}
    u_{(2)}^\mu=v^{\mu\nu}\vartheta_{(2)\nu}
  +\gamma^{\mu\nu}\vartheta_\nu,\\
    u_{(2)}^\mu\vartheta_\mu+u^\mu\vartheta_{(2)\mu}=0.
  \end{gathered}
\end{equation}
The second relation follows from normalization at $\mathcal{O}(\epsilon^2)$. The transverse covector
\begin{equation}
  \vartheta_\mu^\perp\equiv
  p_\perp{}^\nu{}_\mu\vartheta_\nu
  =p_\perp{}^\nu{}_\mu
  \left(h_{\nu\rho}u_{(2)}^\rho+k_{\nu\rho}u^\rho\right)
  \label{eq:sc-transverse-covariant-velocity}
\end{equation}
is fixed by the transverse drift in $u_{(2)}^\mu$ and by any mixed longitudinal--transverse components of $k_{\mu\nu}$. It can enter the leading mixed stress even though the contravariant leading velocity $u^\mu$ is longitudinal.

The vector $u^\mu$ lies in the two-dimensional longitudinal null space, but the geometry does not fix its direction within that null space. For energy density $\mathcal E$, pressure $p$, and enthalpy density $w={\cal E}+p$, use the linear-spin ideal-stress closure
\begin{equation}
  T^{\mu\nu}[\mathcal E,\widehat u;\mu_{\rm s}]
  =w(\mathcal E)\widehat u^\mu\widehat u^\nu
  +p(\mathcal E)g^{\mu\nu}+\mathcal{O}(\varsigma^2).
  \label{eq:linear-spin-ideal-stress-closure}
\end{equation}
The fields $\mathcal E$ and $\widehat u^\mu$ may have induced $\mathcal{O}(\varsigma)$ responses; Eq.~\eqref{eq:linear-spin-ideal-stress-closure} excludes only an explicit term linear in the spin potential. Thermodynamic terms proportional to $\mu_{\rm s}\Omega$ and the associated stress feedback begin at $\mathcal{O}(\varsigma^2)$ and are not part of the coefficient computed here \cite{Florkowski:2024gtr,Drogosz:2024sff,Armas:2026tif}. The two-parameter ordering makes the linear-spin statement independent of the near-horizon weight carried by quadratic spin thermodynamics. Every contribution proportional to $\mu_{\rm s}\Omega$ is $\mathcal{O}(\varsigma^2)$ and therefore vanishes identically in $\partial_\varsigma|_0$. The subsequent $\epsilon$-order is sector dependent. The transverse rotation sector can contribute as early as $\mathcal{O}(\varsigma^2\epsilon^2)=\mathcal{O}(\varsigma^2\lambda)$, while the mixed and longitudinal sectors can begin at higher powers under the parent orthonormal weights of the contraction.
With

\begin{align}
  {\cal E}&={\cal E}_0+\epsilon^2{\cal E}_2+\mathcal{O}(\epsilon^4),\\ p&=p_0+\epsilon^2p_2+\mathcal{O}(\epsilon^4), \\ w&=w_0+\epsilon^2w_2+\mathcal{O}(\epsilon^4),
\end{align}
the mixed stress coefficients are

\begin{align}
  \tau^\mu{}_{\nu} &=w_0u^\mu\vartheta_\nu+p_0\delta^\mu{}_{\nu},\\ \tau_{(2)}^\mu{}_{\nu} &=w_0\left( u_{(2)}^\mu\vartheta_\nu +u^\mu\vartheta_{(2)\nu}\right)\\ &\quad +w_2u^\mu\vartheta_\nu+p_2\delta^\mu{}_{\nu}. \nonumber
\end{align}

In the unscaled tangent basis, $T^{AB}=\mathcal{O}(\epsilon^{-2})$, while $T^{Aa}$ and $T^{ab}$ begin at $\mathcal{O}(1)$. These powers follow from the ideal stress tensor. For the spin current, use the convective form
\begin{equation}
  \begin{gathered}
    S^{\lambda\mu\nu} =\widehat u^\lambda\Omega^{\mu\nu},\\
    \Omega^{\mu\nu}=-\Omega^{\nu\mu}.
  \end{gathered}
\end{equation}
This ansatz fixes the flux direction but does not require the spin density to be spatial in the fluid rest frame. We impose neither the Frenkel condition $\widehat u_\mu\Omega^{\mu\nu}=0$ nor another spin supplementary condition. Phenomenological models may retain independent electric/boost and magnetic/rotation components without imposing the Frenkel constraint \cite{Daher:2024sce}. The independent spin-connection source in Eq.~\eqref{eq:source-currents} couples to all Lorentz components of the canonical spin current. The Ward identities permit both boost- and rotation-sector densities \cite{Hongo:2021ona,Gallegos:2022jow,Chiarini:2024cuv}. Whether a medium retains the boost-sector density is a constitutive choice, not a consequence of the Ward identities or the string-Carroll contraction. The six-component model used here differs from a model that imposes a rest-frame spatial spin tensor.

Equations~\eqref{eq:sc-spin-current-series} and \eqref{eq:sc-velocity-series} fix
\begin{equation}
  \begin{aligned}
    \Omega^{\mu\nu}
    &=\varsigma\epsilon\left(\Sigma^{\mu\nu}
    +\epsilon^2\Sigma_{(2)}^{\mu\nu}+\mathcal{O}(\epsilon^4)\right)\\
    &\quad+\mathcal{O}(\varsigma^2),\\
    \Sigma^{\mu\nu}&=-\Sigma^{\nu\mu}.
  \end{aligned}
  \label{eq:sc-spin-density-scaling}
\end{equation}
Since $\widehat u_\mu=\epsilon\vartheta_\mu+\mathcal{O}(\epsilon^3)$, imposing the Frenkel condition would instead give
\begin{equation}
  \begin{gathered}
    \vartheta_\mu\Sigma^{\mu\nu}=0\\
    \text{at leading order}.
  \label{eq:sc-contracted-frenkel-condition}
  \end{gathered}
\end{equation}
The coefficients of $\Sigma^{\mu\nu}$ in the unscaled contracted frame are not the physical orthonormal components of the parent spin density. Define $\Sigma^{IJ}\equiv e_\mu{}^I e_\nu{}^J\Sigma^{\mu\nu}$ and
\begin{equation}
  \Omega_{\rm phys}^{IJ}\equiv
  \widehat E_\mu{}^I\widehat E_\nu{}^J\Omega^{\mu\nu}.
\end{equation}
Equations~\eqref{eq:rank-two-frame-scaling} and \eqref{eq:sc-spin-density-scaling} then give

\begin{align}
  \Omega_{\rm phys}^{AB}&=\varsigma\epsilon^3\Sigma^{AB}+\mathcal{O}(\varsigma\epsilon^5)+\mathcal{O}(\varsigma^2),\\
  \Omega_{\rm phys}^{Aa}&=\varsigma\epsilon^2\Sigma^{Aa}+\mathcal{O}(\varsigma\epsilon^4)+\mathcal{O}(\varsigma^2),\\
  \Omega_{\rm phys}^{ab}&=\varsigma\epsilon\Sigma^{ab}+\mathcal{O}(\varsigma\epsilon^3)+\mathcal{O}(\varsigma^2).
\end{align}
Longitudinal, mixed, and transverse contracted spin components have different parent-frame weights. The corresponding current coefficients are

\begin{align}
  s^{\lambda\mu\nu} &=u^\lambda\Sigma^{\mu\nu},\\ s_{(2)}^{\lambda\mu\nu} &=u_{(2)}^\lambda\Sigma^{\mu\nu} +u^\lambda\Sigma_{(2)}^{\mu\nu}.
\end{align}

The leading spin flux is longitudinal, and transverse flux first appears at $\mathcal{O}(\epsilon^2)$. Let the parent spin chemical potential have the same contraction weight,
\begin{equation}
\begin{split}
  \mu_{\rm s}^{\mu\nu}
  &=\varsigma\epsilon\varpi^{\mu\nu}
  +\mathcal{O}(\varsigma\epsilon^3)+\mathcal{O}(\varsigma^2),
  \\
  \widehat\Delta^\mu{}_{\nu}
  &=\delta^\mu{}_{\nu}+\widehat u^\mu\widehat u_{\nu}.
  \end{split}
\end{equation}
Its electric/boost vector and magnetic/rotation tensor are

\begin{align}
  &\begin{gathered}
      b^\mu\equiv\mu_{\rm s}^{\mu\nu}\widehat u_\nu,\\
      \widehat u_\mu b^\mu=0,
    \end{gathered}\\
  &\begin{gathered}
      \mu_{\rm R}^{\mu\nu}\equiv
  \widehat\Delta^\mu{}_{\alpha}\widehat\Delta^\nu{}_{\beta}
  \mu_{\rm s}^{\alpha\beta},\\
      \widehat u_\mu\mu_{\rm R}^{\mu\nu}=0.
    \end{gathered}
\end{align}
Thus $\mu_{\rm s}^{\mu\nu}=2\widehat u^{[\mu}b^{\nu]}+\mu_{\rm R}^{\mu\nu}$. We use the sector-dependent nondissipative closure
\begin{equation}
  \begin{gathered}
    \Omega^{\mu\nu}
    =2\chi_{\rm B}({\cal E})\widehat u^{[\mu}b^{\nu]}
    +\chi_{\rm R}({\cal E})\mu_{\rm R}^{\mu\nu},\\
    \chi_{\rm B}<0,\\
    \chi_{\rm R}>0.
  \end{gathered}
  \label{eq:sc-sector-spin-susceptibilities}
\end{equation}
In the local rest frame, $\Omega^{0i}=\chi_{\rm B}\mu_{\rm s}^{0i}$ and $\Omega^{ij}=\chi_{\rm R}\mu_{\rm s}^{ij}$. The signs in Eq.~\eqref{eq:sc-sector-spin-susceptibilities} are necessary low-momentum signs for the boost and rotation sectors in the stated convention \cite{Daher:2024sce}. The Frenkel limit is the boundary value $\chi_{\rm B}=0$; the model studied below instead retains $\chi_{\rm B}<0$ and therefore a nonzero longitudinal boost response. The susceptibilities have dimension $[\Omega]/[\mu_{\rm s}]$ and remain finite in the balanced contraction. The sector split avoids the known one-susceptibility sign conflict but does not establish nonlinear causality, symmetric hyperbolicity, or thermodynamic completeness \cite{Abboud:2025psf}. No derivative constitutive term is included. The subsequent Ward equations use only the sector separation and do not assume a numerical relation between $\chi_{\rm B}$ and $\chi_{\rm R}$.

The fully raised ideal stress is symmetric at both spin orders. This symmetry is a constitutive restriction within the fixed source-defined canonical pseudo-gauge. It is not generated by a Belinfante--Rosenfeld improvement. Consequently, $\tau_{[a]}^\mu{}_{\rho}v^{\rho\nu}$ is symmetric in $\mu,\nu$ for $a=0,1$. On a general branch, the leading spin equation therefore requires
\begin{equation}
  {\cal Q}_{-2}[s]^{\mu\nu}=0. \label{eq:sc-generic-spin-constraint}
\end{equation}
Antisymmetric derivative corrections to the canonical stress tensor lie outside the ideal closure. Define the regular geometric branch by
\begin{equation}
  \Gamma^{(-2)\rho}{}_{\mu\nu}=0. \label{eq:sc-regular-branch}
\end{equation}

Because $\Gamma^{(-2)}$ is a tensor under $\epsilon$-independent coordinate changes that preserve the expansion, this condition is covariant within the contraction. It sets $R^{(-4)}=R^{(-2)}=\mathfrak R^{(-2)}=0$, and Eq.~\eqref{eq:sc-generic-spin-constraint} then holds automatically. Define $\bar\nabla\equiv\nabla^{(0)}$. Expanding the parent metric-compatibility condition gives
\begin{equation}
  \begin{gathered}
    \bar\nabla_\rho h_{\mu\nu}=0,\\
    \bar\nabla_\rho v^{\mu\nu}=0.
  \end{gathered}
\end{equation}

The leading equations for the coefficient linear in $\varsigma$ are
\begin{equation}
  \begin{gathered}
    \bar\nabla_\mu\tau_{[1]}^\mu{}_{\nu}
    ={\cal F}^{(0)}_{[1]\nu},\\
    {\cal F}^{(0)}_{[1]\nu}
    =-\frac12 {\mathfrak R}^{(0)}_{\nu\alpha\beta\gamma}
    u^\alpha\Sigma^{\beta\gamma},\\
    \bar\nabla_\lambda \left(u^\lambda\Sigma^{\mu\nu}\right)=0.
  \end{gathered}
\end{equation}
On this branch, $\mathfrak R^{(0)}_{\nu\alpha\beta\gamma}=h_{\nu\rho}R^{(0)\rho}{}_{\alpha\beta\gamma}$. The physical force is $\varsigma{\cal F}^{(0)}_{[1]\nu}+\mathcal{O}(\varsigma^2)$. The curvature source survives the contraction and is a transverse covector at leading order. The spin equation is equivalently
\begin{equation}
  \begin{gathered}
    u^\lambda\bar\nabla_\lambda\Sigma^{\mu\nu} +\Theta\Sigma^{\mu\nu}=0,\\
    \Theta\equiv\bar\nabla_\lambda u^\lambda.
  \end{gathered}
\end{equation}

Before imposing spacetime symmetry, the longitudinal flow transports all six spin components. Because the longitudinal null space is two-dimensional, the stress equation has energy and momentum projections. Using the covariant longitudinal tensor $v_{\mu\nu}$ from Eq.~\eqref{eq:leading-frame-complements}, define its volume form by $\varepsilon^{(L)}_{\mu\nu}=\varepsilon_{AB}e_\mu{}^A e_\nu{}^B$, with $\varepsilon_{01}=+1$. For a normalized future-directed $u^\mu$, define the orthogonal spacelike vector $z^\mu$ by
\begin{equation}
  \begin{gathered}
    h_{\mu\nu}z^\nu=0,\\
    v_{\mu\nu}z^\mu z^\nu=1,\\
    v_{\mu\nu}u^\mu z^\nu=0,\\
    \varepsilon^{(L)}_{\mu\nu}u^\mu z^\nu=+1.
  \end{gathered}
\end{equation}
The last condition fixes the orientation. Equations~\eqref{eq:sc-leading-covariant-velocity} and \eqref{eq:sc-transverse-covariant-velocity} give $\vartheta_\mu=v_{\mu\nu}u^\nu+\vartheta_\mu^{\perp}$, where $v^{\mu\nu}\vartheta_\nu^{\perp}=0$. The transverse part can enter the leading mixed stress, but it annihilates the longitudinal vectors $u^\mu$ and $z^\mu$, so $\vartheta_\mu z^\mu=0$. The spinless stress $\tau_{[0]}$ obeys the same equations with zero source. Since the linear-spin force has no longitudinal projection, the background and linearized longitudinal equations share the homogeneous form; suppressing $[0]$ on the background fields, they are
\begin{subequations}\label{eq:sc-longitudinal-equations}
  \begin{align}
    u^\mu\partial_\mu{\cal E}_0 +w_0\Theta &=0,\\ z^\nu\partial_\nu p_0 +w_0u^\mu z^\nu\bar\nabla_\mu\vartheta_\nu &=0.
  \end{align}
\end{subequations}

The identity $\vartheta_\nu\bar\nabla_\mu u^\nu=0$ follows from $u^\mu=v^{\mu\nu}\vartheta_\nu$, $v^{\mu\nu}\vartheta_\mu\vartheta_\nu=-1$, and $\bar\nabla_\rho v^{\mu\nu}=0$. The second line is the extra longitudinal momentum equation produced by the two-dimensional longitudinal null space.

On the regular branch, the first post-string-Carroll correction is
\begin{subequations}\label{eq:sc-regular-curvature-correction}
  \begin{align}
    \Gamma^\rho{}_{\mu\nu} &=\Gamma^{(0)\rho}{}_{\mu\nu} +\epsilon^2\Gamma^{(2)\rho}{}_{\mu\nu} +\mathcal{O}(\epsilon^4),\\ R^{(2)\rho}{}_{\sigma\mu\nu} &=2\bar\nabla_{[\mu} \Gamma^{(2)\rho}{}_{\nu]\sigma},\\ {\mathfrak R}^{(2)}_{\nu\alpha\beta\gamma} &=k_{\nu\rho}R^{(0)\rho}{}_{\alpha\beta\gamma} +h_{\nu\rho}R^{(2)\rho}{}_{\alpha\beta\gamma}.
  \end{align}
\end{subequations}

For a mixed tensor $X^\mu{}_{\nu}$ and spin current $Y^{\lambda\mu\nu}$, define

\begin{align}
  \Delta_T[X]_{\nu} &\equiv \Gamma^{(2)\mu}{}_{\mu\rho}X^\rho{}_{\nu} -\Gamma^{(2)\rho}{}_{\mu\nu}X^\mu{}_{\rho},\\ \Delta_S[Y]^{\mu\nu} &\equiv \Gamma^{(2)\lambda}{}_{\lambda\rho}Y^{\rho\mu\nu} +\Gamma^{(2)\mu}{}_{\lambda\rho}Y^{\lambda\rho\nu}\\ &\quad +\Gamma^{(2)\nu}{}_{\lambda\rho}Y^{\lambda\mu\rho}. \nonumber
\end{align}
The $\mathcal{O}(\epsilon^2)$ Ward equations are
\begin{subequations}
  \label{eq:sc-regular-post-system}
  \begin{align}
    \bar\nabla_\mu\tau_{(2)[1]}^\mu{}_{\nu}
    +\Delta_T[\tau_{[1]}]_{\nu}
    &=-\frac12 {\mathfrak R}^{(2)}_{\nu\alpha\beta\gamma}
    s^{\alpha\beta\gamma}\nonumber\\
    &\quad-\frac12 {\mathfrak R}^{(0)}_{\nu\alpha\beta\gamma}
    s_{(2)}^{\alpha\beta\gamma}.
  \end{align}
  \begin{align}
    \bar\nabla_\lambda s_{(2)}^{\lambda\mu\nu}
    +\Delta_S[s]^{\mu\nu}&=0.
  \end{align}
\end{subequations}

The term $k_{\nu\rho}R^{(0)\rho}{}_{\alpha\beta\gamma}$ in Eq.~\eqref{eq:sc-regular-curvature-correction} need not be transverse. The correction may generate a longitudinal projection of the curvature spin force even though $\mathcal F^{(0)}_{[1]\nu}$ is transverse.

The contracted equations are covariant under the string-Carroll boost in Eq.~\eqref{eq:string-carroll-boost}. The projected spin components transform as in Eq.~\eqref{eq:spin-block-boost}. Each order in $\epsilon$ follows from the locally Lorentz-covariant parent identities, so the scaling in Eq.~\eqref{eq:sc-spin-density-scaling} is closed under the boost.

In four dimensions, after specifying $p_0=p_0({\cal E}_0)$ and imposing no spin supplementary condition, the leading ideal system contains one energy density, one longitudinal rapidity, two transverse drift coefficients in $\vartheta_\mu^\perp$, and six components of $\Sigma^{\mu\nu}$. These are ten matter variables. The two drifts are components of $u_{(2)}^\mu$ through Eq.~\eqref{eq:sc-transverse-covariant-velocity}, not independent leading covectors. The contraction hierarchy is triangular and mixes coefficients from different powers of $\epsilon$. The four leading stress equations include the transverse drifts, while the six spin equations transport the spin components. The post-leading system introduces the remaining velocity, thermodynamic, and spin corrections. The algebraic two-susceptibility closure in Eq.~\eqref{eq:sc-sector-spin-susceptibilities} adds no hydrodynamic variable and is not needed to count the transported spin components.

Setting $\Sigma^{\mu\nu}=0$ gives the spinless string-Carroll stress system. Removing one longitudinal direction also removes $z^\mu$, the second momentum equation, the longitudinal rapidity, and $\Sigma^{AB}$. The remaining kinematics matches rank-one Carroll fluids \cite{Freidel:2022bai,Armas:2023dcz,Shukla:2026chs}. The dynamics differ because the canonical equations retain the Riemann spin source.

\section{Static spherical outer horizons in areal-radius gauge}
\label{sec:sss-horizon}

Consider a four-dimensional static, spherically symmetric metric in areal-radius gauge.
\begin{subequations}\label{eq:sss-static-metric}
  \begin{align}
    ds^2&=-e^{2\psi(r)}F(r)dt^2+\frac{dr^2}{F(r)}+r^2d\Omega_2^2,\\ d\Omega_2^2&=d\theta^2+\sin^2\theta\,d\phi^2.
  \end{align}
\end{subequations}

This form includes static spherical black holes with regular nonextremal Killing horizons \cite{Visser:1992qh}. We assume that the areal radius is a smooth local coordinate and is monotone on the exterior side of the future outer horizon $r=r_h$. The functions $F(r)$ and $\psi(r)$ are smooth in this gauge, with $F(r_h)=0$, finite $\psi_h\equiv\psi(r_h)$, and $f_1\equiv F'(r_h)>0$. The last condition fixes the outward orientation and implies nonextremality. Horizons at which the areal radius is not a valid local coordinate, as well as inner and cosmological horizons with different orientation conventions, are outside the present spherical reduction. The line element $d\Omega_2^2$ describes the unit two-sphere.

The static chart is singular at the future horizon. Introduce the ingoing coordinate $v$ through $dv=dt+e^{-\psi(r)}dr/F(r)$. Equation~\eqref{eq:sss-static-metric} then becomes
\begin{equation}
  ds^2=-e^{2\psi(r)}F(r)dv^2 +2e^{\psi(r)}dv\,dr+r^2d\Omega_2^2. \label{eq:sss-ingoing-metric}
\end{equation}

The Killing field is $\chi^\mu=(\partial_v)^\mu$. Define its surface gravity by $\nabla_\mu(\chi^2)=-2\kappa\chi_\mu$ on the horizon. Since $\chi^2=-e^{2\psi}F$ and $\chi_r=e^\psi$ at $r_h$,
\begin{equation}
  \kappa=\frac12 e^{\psi_h}f_1. \label{eq:sss-surface-gravity}
\end{equation}
The branch considered here has $\kappa\neq0$.

Introduce a dimensionless near-horizon parameter $\lambda$ and a local radial coordinate $\rho$ with length dimension through $r=r_h+\lambda\rho$. Define $f_2\equiv F''(r_h)$ and $\psi_1\equiv\psi'(r_h)$. Near the horizon,

\begin{align}
  F&=f_1\lambda\rho+\frac12f_2\lambda^2\rho^2 +\mathcal{O}(\lambda^3),\\ \psi&=\psi_h+\psi_1\lambda\rho+\mathcal{O}(\lambda^2).
\end{align}

Unless explicitly stated otherwise, every $\mathcal{O}(\lambda^n)$ remainder below is defined by the limit $\lambda\to0$ at fixed $\rho$. This limiting prescription is part of the string-Carroll contraction.

Here $[f_1]=[\psi_1]=L^{-1}$, $[f_2]=L^{-2}$, and $[\kappa]=L^{-1}$, while $N_h\equiv e^{\psi_h}$ is dimensionless. Since $dr=\lambda d\rho$, the ingoing metric becomes
\begin{align}
  ds^2={}&r_h^2d\Omega_2^2 +\lambda\Bigl[-N_h^2f_1\rho\,dv^2 +2N_h\,dv\,d\rho\nonumber\\ &\hspace{1.2cm} +2r_h\rho\,d\Omega_2^2\Bigr] \nonumber\\ &+\lambda^2\Bigl[-N_h^2 \left(\frac12f_2+2\psi_1f_1\right) \rho^2dv^2\nonumber\\ &\hspace{1.6cm} +2N_h\psi_1\rho\,dv\,d\rho +\rho^2d\Omega_2^2\Bigr] +\mathcal{O}(\lambda^3).
\end{align}

The two expansion parameters are related by $\lambda=\epsilon^2$. Comparing this metric with Eq.~\eqref{eq:sc-covariant-metric-series} gives

\begin{align}
  h_{\mu\nu}dx^\mu dx^\nu &=r_h^2d\Omega_2^2,\\ k_{\mu\nu}dx^\mu dx^\nu &=-N_h^2f_1\rho\,dv^2+2N_h\,dv\,d\rho +2r_h\rho\,d\Omega_2^2,\\ \ell_{\mu\nu}dx^\mu dx^\nu &=-N_h^2\left(\frac{f_2}{2}+2\psi_1f_1\right) \rho^2dv^2\\ &\quad+2N_h\psi_1\rho\,dv\,d\rho +\rho^2d\Omega_2^2. \nonumber
\end{align}
The tensor $k_{\mu\nu}$ multiplies $\epsilon^2=\lambda$, while $\ell_{\mu\nu}$ multiplies $\epsilon^4=\lambda^2$. Spherical symmetry sets $u_{(2)}^\theta=u_{(2)}^\phi=0$, and the displayed $k_{\mu\nu}$ has no longitudinal--angular components. Equation~\eqref{eq:sc-transverse-covariant-velocity} gives
\begin{equation}
  \begin{gathered}
    \vartheta_\mu^\perp=0\\
    \text{in the spherical sector}.
  \label{eq:sss-transverse-covariant-velocity}
  \end{gathered}
\end{equation}
The reduced leading stress depends only on the longitudinal rapidity.

The rank-two split before expansion is obtained by defining $r_\epsilon\equiv r_h+\epsilon^2\rho$, $N_\epsilon\equiv e^{\psi(r_\epsilon)}$, and $H_\epsilon\equiv F(r_\epsilon)/\epsilon^2$. The unexpanded metric is $g_{\mu\nu}=\epsilon^2V_{\mu\nu}+\Pi_{\mu\nu}$, where
\begin{subequations}\label{eq:sss-exact-rank-two}
  \begin{align}
    V_{\mu\nu}dx^\mu dx^\nu &=-N_\epsilon^2H_\epsilon\,dv^2 +2N_\epsilon\,dv\,d\rho,\\ \Pi_{\mu\nu}dx^\mu dx^\nu &=r_\epsilon^2d\Omega_2^2.
  \end{align}
\end{subequations}

In $(v,\rho)$ coordinates, the longitudinal block has determinant $-N_\epsilon^2$ and remains nondegenerate as $\epsilon\to0$. Its leading inverse components are $v^{vv}=0$, $v^{v\rho}=N_h^{-1}$, and $v^{\rho\rho}=f_1\rho$. The transverse inverse has $h^{\theta\theta}=r_h^{-2}$ and $h^{\phi\phi}=(r_h^2\sin^2\theta)^{-1}$. These tensors satisfy Eq.~\eqref{eq:rank-two-projectors}.

This split also fixes the connection branch. In Eq.~\eqref{eq:sc-connection-exact}, $\mathcal A^\mu{}_{\nu\sigma}$ contracts the inverse longitudinal tensor with derivatives of $\Pi_{\mu\nu}$. At $\epsilon=0$, $\Pi_{(0)}=r_h^2d\Omega_2^2$ has no $v$ or $\rho$ dependence, while $V_{(0)}^{\mu\nu}$ has support only in the $(v,\rho)$ plane. Hence
\begin{equation}
  \begin{gathered}
    {\cal A}_{(0)}^\mu{}_{\nu\sigma}=0,\\
    \Gamma^{(-2)\mu}{}_{\nu\sigma}=0.
  \end{gathered}
\end{equation}
Every horizon in the areal-radius outer-horizon class defined above lies on the regular branch in Eq.~\eqref{eq:sc-regular-branch}.

Let $i,j\in\{v,\rho\}$ label the two-dimensional longitudinal base. The longitudinal part of $k_{\mu\nu}$ defines
\begin{equation}
  q_{ij}dx^i dx^j =-2N_h\kappa\rho\,dv^2+2N_h\,dv\,d\rho, \label{eq:sss-leading-base}
\end{equation}
where Eq.~\eqref{eq:sss-surface-gravity} has been used. Let $\nabla_i^{(q)}$ denote the Levi--Civita derivative of $q_{ij}$. The independent connection coefficients are
\begin{equation}
  \begin{gathered}
    \bar\Gamma^v{}_{vv}=\kappa,\\
    \bar\Gamma^\rho{}_{vv}=2\kappa^2\rho,\\
    \bar\Gamma^\rho{}_{v\rho}=-\kappa, \label{eq:sss-leading-connection}
  \end{gathered}
\end{equation}
and the two-dimensional Riemann tensor of $q_{ij}$ vanishes. The base is ingoing two-dimensional Rindler space, while the leading transverse geometry is a sphere of radius $r_h$. The same split appears in earlier string-Carroll near-horizon formulations \cite{Bagchi:2023cfp,Bagchi:2024rje,Bagchi:2026bnh}.

The notation $q_{ij}\oplus r_h^2d\Omega_2^2$ refers only to the leading metric blocks. The finite connection is not a direct-product connection because $\Gamma^{(0)}={\cal A}_{(2)}+{\cal C}_{(0)}$ retains the first radial change of the sphere. Equation~\eqref{eq:sss-connection-correction} gives this correction.

For the unconstrained spin density used here, spherical symmetry leaves two spin components. One lies in the $(v,\rho)$ plane and the other lies on the sphere. Mixed components vanish because $S^2$ has no nonzero $S\mathcal{O}(3)$-invariant tangent vector. Choose fixed bivectors normalized at the horizon,
\begin{equation}
  \begin{gathered}
    \varepsilon_\parallel^{v\rho}=N_h^{-1},\\
    \varepsilon_\perp^{\theta\phi} =\frac{1}{r_h^2\sin\theta}, \label{eq:sss-invariant-bivectors}
  \end{gathered}
\end{equation}
with all other components fixed by antisymmetry. Away from $\rho=0$, these reference bivectors are not locally normalized. The local bivectors are
\begin{subequations}\label{eq:sss-local-bivectors}
  \begin{align}
    \bar\varepsilon_\parallel^{v\rho}(\lambda)
    &=\frac{1}{N_\epsilon}
    =\frac{1}{N_h}\left(1-\lambda\psi_1\rho\right)+\mathcal{O}(\lambda^2),\\
    \bar\varepsilon_\perp^{\theta\phi}(\lambda)
    &=\frac{1}{r_\epsilon^2\sin\theta}
    =\frac{1}{r_h^2\sin\theta}\left(1-\frac{2\lambda\rho}{r_h}\right)+\mathcal{O}(\lambda^2).
  \end{align}
\end{subequations}
The two contracted frames agree at the horizon. Barred amplitudes introduced below denote coefficients in locally normalized bivectors of $V_{\mu\nu}$ and $\Pi_{\mu\nu}$, while tilded amplitudes denote the exact coefficients in the fixed basis. The unadorned amplitudes $s_X$ are the leading fixed-basis coefficients. These amplitudes are not physical orthonormal components of the parent spin density $\Omega^{\mu\nu}$.

The coefficient of the leading spin density at linear order in $\varsigma$ is
\begin{equation}
  \Sigma^{\mu\nu} =s_\parallel(v,\rho)\varepsilon_\parallel^{\mu\nu} +s_\perp(v,\rho)\varepsilon_\perp^{\mu\nu}. \label{eq:sss-spin-reduction}
\end{equation}
The scalar $s_\parallel$ multiplies the Lorentz-boost generator in the local longitudinal $(01)$ plane, whereas $s_\perp$ multiplies the spatial-rotation generator in the transverse $(23)$ plane. They label the antisymmetric index pair, not the flux index. Their ratio compares two Lorentz sectors, not two spatial spin orientations in the fluid rest frame. Under the convective closure, $s^{\lambda\mu\nu}=u^\lambda\Sigma^{\mu\nu}$, and spherical symmetry gives $u^\theta=u^\phi=0$.

The local channel-ratio relation follows without expanding in $\lambda$. For the unexpanded metric in Eq.~\eqref{eq:sss-exact-rank-two}, the full Levi--Civita connection obeys

\begin{align}
\Gamma^v{}_{iv}+\Gamma^\rho{}_{i\rho}
  &=\partial_i\ln N_\epsilon,\\
  &\begin{gathered}
      \Gamma^\theta{}_{i\theta}+\Gamma^\phi{}_{i\phi}
  =2\partial_i\ln r_\epsilon,\\
      i\in\{v,\rho\}.
    \end{gathered}
\end{align}
Here $i$ is a longitudinal coordinate index. Both right-hand sides vanish for $i=v$ because the background is static. The radial derivatives of $N_\epsilon^{-1}$ and $r_\epsilon^{-2}$ cancel the corresponding connection traces, so
\begin{equation}
  \begin{gathered}
    \widehat u^i\nabla_i\bar\varepsilon_\parallel^{\mu\nu}=0,\\
    \widehat u^i\nabla_i\bar\varepsilon_\perp^{\mu\nu}=0.
  \end{gathered}
\end{equation}
Both locally normalized bivectors are parallel transported along any radial spherical flow. At finite $\epsilon\neq0$, decompose the exact coefficient of the contracted spin density as
\begin{align}
  \Sigma_\epsilon^{\mu\nu}
  &\equiv\left.\epsilon^{-1}
  \frac{\partial\Omega^{\mu\nu}}{\partial\varsigma}
  \right|_{\varsigma=0}\nonumber\\
  &=\bar s_\parallel\bar\varepsilon_\parallel^{\mu\nu}
  +\bar s_\perp\bar\varepsilon_\perp^{\mu\nu}\nonumber\\
  &=\widetilde s_\parallel\varepsilon_\parallel^{\mu\nu}
  +\widetilde s_\perp\varepsilon_\perp^{\mu\nu}.
  \label{eq:sss-exact-spin-decomposition}
\end{align}
The tensor $\Sigma_\epsilon^{\mu\nu}$ is the linear-spin coefficient after the common contraction weight $\epsilon$ is removed. Barred amplitudes multiply the local bivectors, while tilded amplitudes multiply the fixed horizon-normalized bivectors. For a symmetric ideal stress tensor, the exact spin Ward identity and the convective closure give
\begin{equation}
  \begin{gathered}
    \widehat D\bar s_X+\widehat\Theta\bar s_X=0,\\
    X\in\{\parallel,\perp\}.
  \end{gathered}
\end{equation}
Here $\widehat D\equiv\widehat u^i\partial_i$ is the derivative along the full fluid velocity and $\widehat\Theta\equiv\nabla_\lambda\widehat u^\lambda$ is its expansion. Provided $\bar s_\parallel\neq0$, the two equations imply
\begin{equation}
  \widehat D\ln\!\left(\frac{\bar s_\perp}{\bar s_\parallel}\right)=0.
\end{equation}
For a stationary ingoing branch, the locally normalized contracted ratio is
\begin{equation}
  \frac{\bar{\mathcal R}_s(\rho)}{\bar{\mathcal R}_s(0)}=1.
  \label{eq:local-channel-ratio}
\end{equation}
Here $\bar{\mathcal R}_s\equiv\bar s_\perp/\bar s_\parallel$. This equality is exact within the ideal radial spherical sector. Derivative spin currents, mixed spin components, nonspherical flow, torsion, or another constitutive closure can modify it. Comparing the two exact decompositions in Eq.~\eqref{eq:sss-exact-spin-decomposition} gives
\begin{equation}
  \begin{gathered}
    \widetilde s_\parallel=\frac{N_h}{N_\epsilon}\bar s_\parallel,\\
    \widetilde s_\perp=\left(\frac{r_h}{r_\epsilon}\right)^2\bar s_\perp.
  \end{gathered}
\end{equation}
The fixed horizon-normalized ratio obeys
\begin{equation}
  \frac{\mathcal R_s^{(h)}(\rho)}{\mathcal R_s^{(h)}(0)}
  =\frac{N_\epsilon}{N_h}
  \left(\frac{r_h}{r_\epsilon}\right)^2.
  \label{eq:exact-fixed-channel-ratio}
\end{equation}
Here $\mathcal R_s^{(h)}\equiv \widetilde s_\perp/\widetilde s_\parallel$. The right-hand side is the geometric conversion between the two contracted bivector bases.
It contains no independent constitutive coefficient and cannot by itself represent spin transport. The advected local ratio in Eq.~\eqref{eq:local-channel-ratio} is the dynamical statement; Eq.~\eqref{eq:exact-fixed-channel-ratio} only reports how its components change when the reference bivectors are kept normalized at the horizon rather than at the observation point.

In this sector, a Frenkel condition removes the longitudinal component. Equation~\eqref{eq:sss-transverse-covariant-velocity} gives $\vartheta_i=u_i=q_{ij}u^j$. Hence the leading Frenkel condition in Eq.~\eqref{eq:sc-contracted-frenkel-condition} would require
\begin{equation}
  \begin{gathered}
    \vartheta_\mu\Sigma^{\mu j}
  =s_\parallel u_i\varepsilon_\parallel^{ij}
  =\pm s_\parallel z^j=0,\\
    \Longrightarrow\quad  s_\parallel=0.
  \label{eq:sss-frenkel-removes-longitudinal-spin}
  \end{gathered}
\end{equation}
The sign depends on the orientation convention. In the sector-dependent closure of Eq.~\eqref{eq:sc-sector-spin-susceptibilities}, the strict Frenkel limit is $\chi_{\rm B}=0$. It removes the longitudinal response while retaining the transverse rotation sector. The model studied below keeps $\chi_{\rm B}<0$, so the boost-sector amplitude can be nonzero without assigning the same susceptibility to both Lorentz sectors.

One rapidity describes the leading fluid velocity. Define two null vectors of $q_{ij}$ by $k_+^i\partial_i=\partial_v+\kappa\rho\,\partial_\rho$ and $k_-^i\partial_i=-N_h^{-1}\partial_\rho$. They satisfy $q_{ij}k_+^ik_+^j=q_{ij}k_-^ik_-^j=0$ and $q_{ij}k_+^ik_-^j=-1$. For a dimensionless rapidity $\eta(v,\rho)$, set
\begin{equation}
  \begin{gathered}
    u^i=\frac{e^\eta k_+^i+e^{-\eta}k_-^i}{\sqrt2},\\
    z^i=\frac{e^\eta k_+^i-e^{-\eta}k_-^i}{\sqrt2}. \label{eq:sss-rapidity-frame}
  \end{gathered}
\end{equation}

Then $q_{ij}u^iu^j=-1$, $q_{ij}z^iz^j=1$, and $q_{ij}u^iz^j=0$, so $z^i$ is the spacelike vector used in Eq.~\eqref{eq:sc-longitudinal-equations}. Define the base connection one-form by $\nabla_i^{(q)} k_+^j=\omega_i k_+^j$ and $\nabla_i^{(q)} k_-^j=-\omega_i k_-^j$. Equation~\eqref{eq:sss-leading-connection} gives $\omega_i dx^i=\kappa\,dv$. With $D\equiv u^i\partial_i$, the expansion $\Theta$ and longitudinal acceleration $a_\parallel$ are
\begin{equation}
  \begin{gathered}
    \Theta=z^i(\partial_i\eta+\omega_i),\\
    a_\parallel=u^i(\partial_i\eta+\omega_i).
  \end{gathered}
\end{equation}

After the spherical reduction, the $\mathcal{O}(\lambda^0)$ curvature spin force vanishes. The leading stress and spin Ward identities then form a closed $1+1$ system. For $w_0={\cal E}_0+p_0$ and $p_0=p_0({\cal E}_0)$, the independent equations are
\begin{equation}
  \begin{aligned} D{\cal E}_0+w_0\Theta&=0,\\ w_0a_\parallel+z^i\partial_i p_0&=0,\\ Ds_\parallel+\Theta s_\parallel&=0,\\ Ds_\perp+\Theta s_\perp&=0.
  \end{aligned} \label{eq:sss-leading-reduced-system}
\end{equation}

The first two lines are the energy and longitudinal momentum equations. The last two follow from $\bar\nabla_\lambda(u^\lambda\Sigma^{\mu\nu})=0$ because the bivectors in Eq.~\eqref{eq:sss-invariant-bivectors} are parallel along the leading longitudinal flow. Their identical advection--dilution form does not assume equal boost- and rotation-sector susceptibilities. The angular stress equations hold identically by $S\mathcal{O}(3)$ symmetry.

Using the notation of Eq.~\eqref{eq:sc-regular-curvature-correction}, write the first connection correction as $\Gamma^\mu{}_{\nu\sigma}=\bar\Gamma^\mu{}_{\nu\sigma}+\lambda\Gamma^{(2)\mu}{}_{\nu\sigma}+\mathcal{O}(\lambda^2)$. The components needed for the radial equations are
\begin{subequations}\label{eq:sss-connection-correction}
  \begin{align}
    \Gamma^{(2)v}{}_{vv} &=\frac{N_h}{2}(f_2+3f_1\psi_1)\rho,\\ \Gamma^{(2)\rho}{}_{v\rho} &=-\Gamma^{(2)v}{}_{vv},\\ \Gamma^{(2)\rho}{}_{vv} &=\frac{N_h^2f_1}{4} (3f_2+8f_1\psi_1)\rho^2,\\ \Gamma^{(2)\rho}{}_{\rho\rho}&=\psi_1,\\ \Gamma^{(2)\theta}{}_{\rho\theta} &=\frac1{r_h},\\ \Gamma^{(2)\phi}{}_{\rho\phi} &=\frac1{r_h}.
  \end{align}
\end{subequations}

Define $C_i\equiv\Gamma^{(2)\mu}{}_{\mu i}$. Equation~\eqref{eq:sss-connection-correction} gives $C_v=0$ and $C_\rho=\psi_1+2/r_h$. The $\mathcal{O}(\lambda)$ geometry depends on $f_2$ and $\psi_1$.

For the spherical ideal stress tensor, the $\mathcal{O}(\lambda)$ curvature spin terms in Eq.~\eqref{eq:sc-regular-post-system} vanish. Let $\tau_{(2)}^i{}_j$ be the $\mathcal{O}(\lambda)$ part of the mixed stress tensor and set $u_i=q_{ij}u^j$. For $i\in\{v,\rho\}$, the radial stress equations reduce to
\begin{align}
  \nabla_j^{(q)}\tau_{(2)}^j{}_i +w_0 C_j u^j u_i -w_0\Gamma^{(2)k}{}_{ji}u^j u_k&=0. \label{eq:sss-post-stress}
\end{align}
The pressure terms cancel between the base and angular directions. Equation~\eqref{eq:sss-post-stress} contains the geometric source for the ideal radial stress equations at this order.

Spherical symmetry restricts $\Sigma_{(2)}^{\mu\nu}$ to the same two bivectors. Write $\Sigma_{(2)}^{\mu\nu}=s_{\parallel(2)}\varepsilon_\parallel^{\mu\nu}+s_{\perp(2)}\varepsilon_\perp^{\mu\nu}$ and define the first-order flux corrections by
\begin{equation}
  \begin{gathered}
    J_{\parallel(2)}^i \equiv u_{(2)}^i s_\parallel+u^i s_{\parallel(2)},\\
    J_{\perp(2)}^i \equiv u_{(2)}^i s_\perp+u^i s_{\perp(2)}. \label{eq:sss-spin-correction-fluxes}
  \end{gathered}
\end{equation}
Here $u_{(2)}^i$ is the $\mathcal{O}(\lambda)$ longitudinal velocity in Eq.~\eqref{eq:sc-velocity-series}. The spin equations at this order are
\begin{subequations}\label{eq:sss-post-spin}
  \begin{align}
    \nabla_i^{(q)}J_{\parallel(2)}^i +2\left(\psi_1+\frac1{r_h}\right)u^\rho s_\parallel&=0,\\ \nabla_i^{(q)}J_{\perp(2)}^i +\left(\psi_1+\frac4{r_h}\right)u^\rho s_\perp&=0.
  \end{align}
\end{subequations}
The two channels receive different geometric corrections. The $f_2$ terms cancel from Eq.~\eqref{eq:sss-post-spin}, but remain in the radial stress correction through Eq.~\eqref{eq:sss-connection-correction}.

For the unexpanded ingoing metric, the independent Riemann component of the $(v,r)$ base is
\begin{equation}
  R_{vrvr}=\frac{e^{2\psi}}{2} \left[F''+3\psi'F' +2F\left(\psi''+(\psi')^2\right)\right].
\end{equation}

Under $r=r_h+\lambda\rho$, each covariant $\rho$ index contributes one factor of $\lambda$, so $R_{v\rho v\rho}=\lambda^2R_{vrvr}$. Define $\mathcal K_h\equiv f_2+3f_1\psi_1$, with dimension $L^{-2}$. Denote by $\mathcal F_\nu$ the coefficient of the force linear in $\varsigma$, so the physical force is $\varsigma\mathcal F_\nu+\mathcal{O}(\varsigma^2)$. Substituting Eq.~\eqref{eq:sss-spin-reduction} gives
\begin{subequations}\label{eq:sss-first-spin-force}
  \begin{align}
\mathcal F_v &=-\frac{\lambda^2N_h}{2}\mathcal K_h u^\rho s_\parallel+\mathcal{O}(\lambda^3),\\
  &\begin{gathered}
      \mathcal F_\rho =\frac{\lambda^2N_h}{2}\mathcal K_h u^v s_\parallel+\mathcal{O}(\lambda^3),\\
      \mathcal F_\theta=\mathcal F_\phi=0.
    \end{gathered}
\end{align}
\end{subequations}

The combination $\mathcal K_h$ has an invariant geometric meaning. On the unscaled $(v,r)$ base, let $\varepsilon_{(L)}^{\mu\nu}$ be its unit volume bivector, with $\varepsilon_{(L)}^{vr}=e^{-\psi(r)}$. Define the horizon longitudinal curvature scalar by
\begin{equation}
  \mathcal R^{(L)}_h\equiv
  \left.\frac14R_{\mu\nu\alpha\beta}
  \varepsilon_{(L)}^{\mu\nu}
  \varepsilon_{(L)}^{\alpha\beta}\right|_{r_h}
  =\frac12\mathcal K_h.
  \label{eq:sss-longitudinal-curvature-scalar}
\end{equation}
This scalar is invariant under regular reparametrizations of the $(v,r)$ base and under local $S\mathcal{O}(1,1)$ frame changes. For Reissner--Nordstr\"om, $-\mathcal R^{(L)}_h$ is the radial geodesic-deviation eigenvalue in the convention used for tidal acceleration. Its zero is therefore the previously known horizon tidal inversion \cite{Crispino:2016rnm}.

The transverse rotation-sector amplitude $s_\perp$ does not enter the force. The warped-product Riemann tensor has no component with one radial-flow index and an antisymmetric pair entirely on the sphere, and $S\mathcal{O}(3)$ forbids mixed spin bivectors. In the unconstrained-spin model, the spherical curvature coupling acts through the longitudinal boost-sector channel. Its first physical contribution is $\mathcal{O}(\varsigma\lambda^2)=\mathcal{O}(\varsigma\epsilon^4)$ and is fixed by $\mathcal K_h$. A Frenkel-constrained model has no such contribution because $s_\parallel=0$.

Setting $s_\parallel=s_\perp=0$ gives the spinless ideal string-Carroll system. If $f_2=\psi_1=0$ and the sphere is replaced locally by its planar large-radius limit, all correction sources vanish and Eq.~\eqref{eq:sss-leading-reduced-system} becomes the flat Rindler system. The derivation assumes $\kappa\neq0$. Setting $\kappa=0$ does not produce an extremal horizon. Regular extremal near-horizon geometries require a separate scaling, and in RN the nominally subleading terms become essential as extremality is approached \cite{Kunduri:2013gce,Shinde:2026ern}.

\section{Analytic near-horizon spin transport}
\label{sec:analytic-spin-transport}

Without assuming stationarity, the reduced spin equations integrate once. For a barotropic equation of state $p_0=p_0({\cal E}_0)$, define
\begin{equation}
  \begin{gathered}
    {\cal N}({\cal E}_0) \equiv \exp\left[ \int_{{\cal E}_h}^{{\cal E}_0} \frac{d{\cal E}}{w_0({\cal E})}\right],\\
    w_0({\cal E})\equiv{\cal E}+p_0({\cal E}), \label{eq:transport-density-factor}
  \end{gathered}
\end{equation}
where $\mathcal E_h$ is the energy density at which a fluid trajectory crosses the future horizon and $w_0>0$ on the chosen branch. The energy and spin equations in Eq.~\eqref{eq:sss-leading-reduced-system} give
\begin{equation}
  \begin{gathered}
    D\left(\frac{s_\parallel}{{\cal N}}\right)=0,\\
    D\left(\frac{s_\perp}{{\cal N}}\right)=0,\\
    D\left(\frac{s_\perp}{s_\parallel}\right)=0,
  \end{gathered}
\end{equation}
where the last relation requires $s_\parallel\neq0$. Both amplitudes have the same dilution factor, so their ratio is constant along the flow. The same definition gives $D{\cal N}+\Theta{\cal N}=0$.

For stationary radial flow, the reduced system admits a regular solution. Stationarity means $\partial_v{\cal E}_0=\partial_v\eta=\partial_v s_\parallel=\partial_v s_\perp=0$ in the ingoing coordinates of Eq.~\eqref{eq:sss-ingoing-metric}. Since the longitudinal metric in Eq.~\eqref{eq:sss-leading-base} has $\det q=-N_h^2$, the density, spin, and Killing-energy equations give

\begin{align}
  j_{\cal N}\equiv u^\rho{\cal N}&=\text{constant},\\ j_X\equiv u^\rho s_X&=\text{constant},\\ {\cal J}_E\equiv\tau^\rho{}_v &=w_0u^\rho u_v=\text{constant},
\end{align}
where $X\in\{\parallel,\perp\}$ labels the channel and $u_v\equiv q_{vi}u^i$. At the horizon, a finite $\eta_h\equiv\eta(0)$ gives
\begin{equation}
  \begin{gathered}
    u_h^v=\frac{e^{\eta_h}}{\sqrt2},\\
    u_h^\rho=-\frac{e^{-\eta_h}}{\sqrt2\,N_h}.
  \end{gathered}
\end{equation}

Both components are finite, and $u_h^\rho<0$ for a future-directed ingoing flow. Regularity requires finite $\mathcal E_h$, $w_h\equiv w_0({\cal E}_h)>0$, $\eta_h$, $s_{\parallel h}$, and $s_{\perp h}$. It does not relate the two horizon spin amplitudes.

The stationary first integrals can first be reduced for an arbitrary barotrope. Define
\begin{equation}
  \begin{gathered}
    y\equiv N_h\kappa\rho\,e^{2\eta},\\
    x\equiv N_h\kappa\rho\,e^{2\eta_h}.
  \end{gathered}
\end{equation}
Equation~\eqref{eq:sss-rapidity-frame} gives
\begin{subequations}\label{eq:stationary-velocity-y}
  \begin{align}
  &\begin{gathered}
      u^v=\frac{e^\eta}{\sqrt2},\\
      u^\rho=\frac{e^{-\eta}}{\sqrt2\,N_h}(y-1),
    \end{gathered}\\
u_v&=-\frac{e^{-\eta}}{\sqrt2}(y+1).
\end{align}
\end{subequations}

Let
\begin{align}
  R_{\cal N}({\cal E})&\equiv
  \frac{{\cal N}({\cal E})}{{\cal N}({\cal E}_h)},\\
  R_w({\cal E})&\equiv\frac{w_0({\cal E})}{w_h},\\
  H({\cal E})&\equiv
  \frac{R_w({\cal E})}{R_{\cal N}({\cal E})^2}.
\end{align}
Ratios of the density and Killing-energy fluxes to their horizon values give
\begin{equation}
  \begin{gathered}
    R_{\cal N}e^{-(\eta-\eta_h)}(1-y)=1,\\
    R_we^{-2(\eta-\eta_h)}(1-y^2)=1.
  \end{gathered}
\end{equation}
Consequently, on any interval on which $H$ is one-to-one, the leading stationary solution is the analytic reduction
\begin{subequations}\label{eq:stationary-general-barotrope-solution}
  \begin{align}
{\cal E}_0(y)
    &=H^{-1}\!\left(\frac{1-y}{1+y}\right),\\
  &\begin{gathered}
      \frac{s_X(y)}{s_{Xh}}
    =R_{\cal N}({\cal E}_0(y)),\\
      X\in\{\parallel,\perp\},
    \end{gathered}\\
e^{\eta-\eta_h}
    &=R_{\cal N}({\cal E}_0(y))(1-y),\\
x&=\frac{y}{R_{\cal N}({\cal E}_0(y))^2(1-y)^2}.
\end{align}
\end{subequations}
Equation~\eqref{eq:stationary-general-barotrope-solution} uses one thermodynamic quadrature, Eq.~\eqref{eq:transport-density-factor}, followed by one local analytic inverse. No radial integration or numerical evolution is required.

An elementary closed form follows for the affine constant-sound-speed family
\begin{equation}
  \begin{gathered}
    p_0=\alpha{\cal E}_0+\Pi,\\
    0\leq\alpha<1,\\
    w_0=(1+\alpha){\cal E}_0+\Pi>0,
  \label{eq:affine-barotrope}
  \end{gathered}
\end{equation}
where $\Pi$ is constant and $\alpha=dp_0/d{\cal E}_0$ is the squared sound speed. This strictly contains the proportional barotrope as $\Pi=0$. For $R_w\equiv w_0/w_h$, Eq.~\eqref{eq:transport-density-factor} gives $R_{\cal N}=R_w^{1/(1+\alpha)}$. Solving the two first integrals algebraically gives the ingoing branch connected continuously to $y=0$:
\begin{subequations}\label{eq:stationary-parametric-solution}
  \begin{align}
R_w
    &=\left(\frac{1+y}{1-y}\right)^{(1+\alpha)/(1-\alpha)},\\
  &\begin{gathered}
      {\cal E}_0
    =\frac{w_hR_w-\Pi}{1+\alpha},\\
      p_0=\frac{\alpha w_hR_w+\Pi}{1+\alpha},
    \end{gathered}\\
  &\begin{gathered}
      \frac{s_X}{s_{Xh}}
    =\left(\frac{1+y}{1-y}\right)^{1/(1-\alpha)},\\
      X\in\{\parallel,\perp\},
    \end{gathered}\\
e^{\eta-\eta_h}
    &=(1+y)^{1/(1-\alpha)}(1-y)^{-\alpha/(1-\alpha)},\\
x&=y(1-y)^{2\alpha/(1-\alpha)}(1+y)^{-2/(1-\alpha)}.
\end{align}
\end{subequations}

This is an explicit closed-form parametric solution: $y$ is the parameter, and every field and the corresponding radius $x$ are elementary functions of it. It satisfies the leading reduced equations without radial integration. For pressureless matter, $\alpha=0$ and $\Pi=0$, it reduces to
\begin{subequations}\label{eq:stationary-dust-limit}
  \begin{align}
\frac{{\cal E}_0}{{\cal E}_h}
    =\frac{s_X}{s_{Xh}}
    &=\frac{1+y}{1-y},\\
  &\begin{gathered}
      e^{\eta-\eta_h}=1+y,\\
      x=\frac{y}{(1+y)^2}.
    \end{gathered}
\end{align}
\end{subequations}
On the branch continuous at the horizon, the last relation can be inverted in elementary form,
\begin{equation}
  \begin{gathered}
    y(x)=\frac{1-2x-\sqrt{1-4x}}{2x},\\
    0<x<\frac14,\\
    y(0)=0.
  \end{gathered}
\end{equation}
Thus the dust fields are also explicit functions of radius rather than only parametric functions. These relations follow directly from the stationary dust equations.

The stationary equations restrict the domain of this solution. Define $\Delta_{\rm s}\equiv(u^\rho)^2-\alpha(z^\rho)^2$, where $z^\rho$ is the radial component of the unit spacelike vector in Eq.~\eqref{eq:sss-rapidity-frame}. Equation~\eqref{eq:stationary-velocity-y} gives
\begin{equation}
  \Delta_{\rm s} =\frac{e^{-2\eta}}{2N_h^2} \left[(1-y)^2-\alpha(1+y)^2\right]. \label{eq:stationary-sonic-denominator}
\end{equation}

The determinant of the stationary equations for $({\cal E}_0',\eta')$, with a prime denoting $d/d\rho$, is $w_0\Delta_{\rm s}$. Its zero defines a sonic point. For $0<\alpha<1$, the first zero outside the horizon is
\begin{equation}
  y_{\rm s}=\frac{1-\sqrt{\alpha}}{1+\sqrt{\alpha}}.
\end{equation}

The regular near-horizon branch has $0\leq y<y_{\rm s}$. For $\alpha=0$, the endpoint is $y_{\rm s}=1$. Continuing through $y_{\rm s}$ requires a separate sonic matching analysis. Since $y=x+\mathcal{O}(x^2)$ near the horizon, Eq.~\eqref{eq:stationary-parametric-solution} gives

\begin{align}
  R_w&=1+\frac{2(1+\alpha)}{1-\alpha}x+\mathcal{O}(x^2),\\
  {\cal E}_0-{\cal E}_h
  &=\frac{2w_h}{1-\alpha}x+\mathcal{O}(x^2),\\
  \frac{s_X}{s_{Xh}}&=1+\frac{2}{1-\alpha}x+\mathcal{O}(x^2),\\
  \eta-\eta_h&=\frac{1+\alpha}{1-\alpha}x+\mathcal{O}(x^2).
\end{align}

The horizon values and first derivatives are finite for $0\leq\alpha<1$. The stiff case $\alpha=1$ is singular in this expansion and requires a different local scaling.

We now compute the stationary response at linear order in $\varsigma$ and at $\mathcal{O}(\lambda^2)$. For the static Killing vector $\chi^\mu=(\partial_v)^\mu$, Eq.~\eqref{eq:killing-spin-current} gives the radial closed-sphere flux
\begin{equation}
  \begin{aligned}
    \Phi_\chi(\rho)
    &\equiv\int_{S^2_\rho}d\theta\,d\phi\,\sqrt{-g}\,
    \left(T^\rho{}_{v}
    +\frac12S^{\rho\alpha\beta}\nabla_\alpha\chi_\beta\right),\\
    \partial_\rho\Phi_\chi&=0.
  \end{aligned}
  \label{eq:stationary-invariant-killing-flux}
\end{equation}
Let the unexpanded barotropic density factor be
\begin{equation}
  \widehat{\cal N}({\cal E})\equiv
  \exp\!\left[\int_{{\cal E}_h}^{\cal E}
  \frac{d\widetilde{\cal E}}{w(\widetilde{\cal E})}\right].
\end{equation}
The remaining exact radial fluxes may be written as
\begin{align}
\Phi_{\cal N}
  &=\int_{S^2_\rho}d\theta\,d\phi\,\sqrt{-g}\,
  \widehat u^\rho\widehat{\cal N},\\
  &\begin{gathered}
      \Phi_X
  =\int_{S^2_\rho}d\theta\,d\phi\,\sqrt{-g}\,
  \widehat u^\rho\bar s_X,\\
      X\in\{\parallel,\perp\},
    \end{gathered}
\end{align}
with $\partial_\rho\Phi_{\cal N}=\partial_\rho\Phi_X=0$. The two terms in parentheses in Eq.~\eqref{eq:stationary-invariant-killing-flux} are separately pseudo-gauge dependent. Equation~\eqref{eq:pseudogauge-killing-superpotential} proves that their sum has an invariant closed-sphere flux for stationary, regular, single-valued improvements. These four fluxes are exact first integrals of the stationary spherical system and check every order of the near-horizon series.

To isolate the curvature--spin response, compare solutions on the same metric and with the same equation of state, $\mathcal E_h$, and $\eta_h$. Define the matched difference
\begin{equation}
  \Delta_{\rm sp}X\equiv X(\epsilon,\varsigma)-X(\epsilon,0)
  =\varsigma X_{[1]}+\mathcal{O}(\varsigma^2).
\end{equation}
All purely geometric post-Carroll terms cancel in $\Delta_{\rm sp}$. Because the force already carries $\lambda^2$, its coefficient linear in $\varsigma$ is evaluated on the spinless leading flow. With $j_\parallel=u^\rho s_\parallel$ the conserved linear-spin coefficient and $\Phi_{E,h}\equiv\tau^\rho{}_v(0)=w_hu_h^\rho u_{vh}$, the $v$ component of the Ward identity gives
\begin{equation}
  \partial_\rho\,\Delta_{\rm sp}\tau^\rho{}_v
  =-\varsigma\frac{\lambda^2N_h}{2}\mathcal K_hj_\parallel
  +\mathcal{O}(\varsigma\lambda^3)+\mathcal{O}(\varsigma^2).
  \label{eq:stationary-matched-energy-flux-equation}
\end{equation}
The leading spin equation makes $j_\parallel$ constant. Integration with matched horizon values yields
\begin{subequations}\label{eq:stationary-spin-response-factor}
  \begin{align}
1+\frac{\Delta_{\rm sp}\tau^\rho{}_v(\rho)}{\Phi_{E,h}}
    &=\mathcal G_{\rm sp}(\rho)
    +\mathcal{O}(\varsigma\lambda^3)+\mathcal{O}(\varsigma^2),\\
  &\begin{gathered}
      \mathcal G_{\rm sp}(\rho)
    \equiv1+\varsigma g(\rho),\\
      g(\rho)\equiv-
    \frac{\lambda^2N_h\mathcal K_hj_\parallel}{2\Phi_{E,h}}\,\rho.
    \end{gathered}
\end{align}
\end{subequations}
The spin term in Eq.~\eqref{eq:stationary-invariant-killing-flux} supplies the complementary radial change, so the total flux $\Phi_\chi$ remains constant. This cancellation is the pseudo-gauge-independent content of the local canonical force.

For a general barotrope, the density-flux relation is unchanged while the matter energy-flux relation acquires $\mathcal G_{\rm sp}$. The analytic reduction in Eq.~\eqref{eq:stationary-general-barotrope-solution} becomes
\begin{subequations}\label{eq:stationary-general-response-solution}
  \begin{align}
    {\cal E}(y,\varsigma)
    &=H^{-1}\!\left(
    \mathcal G_{\rm sp}\frac{1-y}{1+y}\right),\\
    e^{\eta-\eta_h}
    &=R_{\cal N}({\cal E})(1-y),\\
    x&=\frac{y}{R_{\cal N}({\cal E})^2(1-y)^2}.
  \end{align}
\end{subequations}
For the affine equation of state, this gives the explicit response-dressed matter solution
\begin{subequations}\label{eq:stationary-response-dressed-solution}
  \begin{align}
R_w
    &=\left[\frac{1+y}{(1-y)\mathcal G_{\rm sp}}\right]^{(1+\alpha)/(1-\alpha)},\\
  &\begin{gathered}
      {\cal E}
    =\frac{w_hR_w-\Pi}{1+\alpha},\\
      p=\frac{\alpha w_hR_w+\Pi}{1+\alpha},
    \end{gathered}\\
e^{\eta-\eta_h}
    &=(1+y)^{1/(1-\alpha)}
    (1-y)^{-\alpha/(1-\alpha)}
    \mathcal G_{\rm sp}^{-1/(1-\alpha)},\\
x
    &=y(1-y)^{2\alpha/(1-\alpha)}
    (1+y)^{-2/(1-\alpha)}
    \mathcal G_{\rm sp}^{2/(1-\alpha)}.
\end{align}
\end{subequations}
Equations~\eqref{eq:stationary-general-response-solution} and \eqref{eq:stationary-response-dressed-solution} are generating forms that must be Taylor expanded to first order in $\varsigma$ and retained through $\mathcal{O}(\lambda^2)$. Terms quadratic in $\mathcal G_{\rm sp}-1$ are $\mathcal{O}(\varsigma^2)$ and are excluded together with quadratic spin thermodynamics. The domain requires $|\varsigma g|\ll1$ and separation from the sonic point.

The fixed-radius response is fully explicit. Write $y=y_0+\varsigma y_{[1]}+\mathcal{O}(\varsigma^2)$, where $y_0$ is the parameter of the spinless solution at the chosen $x$, and define
\begin{equation}
  D_0(y_0)\equiv(1-y_0)^2-\alpha(1+y_0)^2.
\end{equation}
Expanding the last line of Eq.~\eqref{eq:stationary-response-dressed-solution} at fixed $x$ gives
\begin{subequations}\label{eq:fixed-radius-linear-spin-response}
  \begin{align}
y_{[1]}
    &=-\frac{2g\,y_0(1-y_0^2)}{D_0},\\
\frac{w_{[1]}}{w^{(0)}}
    &=-\frac{(1+\alpha)g}{1-\alpha}
    \left(1+\frac{4y_0}{D_0}\right),\\
\eta_{[1]}
    &=-\frac{g}{1-\alpha}
    \left[1+\frac{2y_0\{(1+\alpha)+(\alpha-1)y_0\}}{D_0}\right],\\
  &\begin{gathered}
      {\cal E}_{[1]}=\frac{w_{[1]}}{1+\alpha},\\
      p_{[1]}=\alpha{\cal E}_{[1]}.
    \end{gathered}
\end{align}
\end{subequations}
These rational expressions provide an analytic approximation without numerical root finding. Their pole at $D_0=0$ is the expected failure of a regular perturbation at the sonic layer. The contracted spin-current amplitudes themselves are, consistently at first order,
\begin{equation}
  \begin{aligned}
    \varsigma s_X(x)
    &=\varsigma s_{Xh}
    \left(\frac{1+y_0}{1-y_0}\right)^{1/(1-\alpha)}
    +\mathcal{O}(\varsigma^2),\\
    &\hspace{5.2em}X\in\{\parallel,\perp\}.
  \end{aligned}
  \label{eq:linear-contracted-spin-profiles}
\end{equation}
The $\mathcal{O}(\varsigma)$ change of the coefficient $s_X$ would multiply the overall $\varsigma$ in $S^{\lambda\mu\nu}$ and therefore contribute only at $\mathcal{O}(\varsigma^2)$, where the omitted thermodynamic feedback must also be restored. We do not include that uncontrolled partial second-order effect.

The local response can also be stated without coordinates. Define $\ell_s=s_\parallel/w_0$ on the spinless leading flow and subtract its Euler equation. The complete linear-spin longitudinal combination within the stated closure is
\begin{equation}
  \begin{aligned}
    \mathcal A_{\rm sp}
    &\equiv\Delta_{\rm sp}\!\left(
    a_\parallel+\frac{z^i\partial_i p_0}{w_0}\right)
    \\
    &=\varsigma\lambda^2\mathcal R^{(L)}_h\ell_s
    +\mathcal{O}(\varsigma\lambda^3)+\mathcal{O}(\varsigma^2)\\
    &=\frac{\varsigma\lambda^2}{2}\mathcal K_h\ell_s
    +\mathcal{O}(\varsigma\lambda^3)+\mathcal{O}(\varsigma^2).
  \end{aligned}
  \label{eq:invariant-spin-response-scalar}
\end{equation}
$\mathcal A_{\rm sp}$ is a scalar under regular base-coordinate changes and local longitudinal-frame transformations. For dust, $p_0=0$, it reduces directly to $\Delta_{\rm sp}a_\parallel=\varsigma\lambda^2\mathcal K_h\ell_s/2+\mathcal{O}(\varsigma\lambda^3)+\mathcal{O}(\varsigma^2)$. For nonzero sound speed, Eq.~\eqref{eq:fixed-radius-linear-spin-response} supplies the pressure and velocity response rather than identifying the force alone with acceleration.

The first post-string-Carroll spin equations also integrate on the stationary branch. Since $\sqrt{-q}=N_h$ is constant and $j_X=u^\rho s_X$ does not depend on $\rho$, Eq.~\eqref{eq:sss-post-spin} gives

\begin{align}
  J_{\parallel(2)}^\rho(\rho) &=J_{\parallel(2)}^\rho(0) -2\left(\psi_1+\frac1{r_h}\right)j_\parallel\rho,\\ J_{\perp(2)}^\rho(\rho) &=J_{\perp(2)}^\rho(0) -\left(\psi_1+\frac4{r_h}\right)j_\perp\rho.
\end{align}

Expand the exact fixed-basis amplitudes as $\widetilde s_X=s_X+\lambda s_{X(2)}+\mathcal{O}(\lambda^2)$. Equation~\eqref{eq:sss-spin-correction-fluxes} gives $J_{X(2)}^\rho/j_X=u_{(2)}^\rho/u^\rho+s_{X(2)}/s_X$. The common velocity term cancels from the ratio. The flux calculation reproduces the Taylor expansion of the exact conversion in Eq.~\eqref{eq:exact-fixed-channel-ratio}.
\begin{align}
  \frac{\mathcal R_s^{(h)}(\rho)}{\mathcal R_s^{(h)}(0)}
  &=\frac{N_\epsilon}{N_h}\left(\frac{r_h}{r_\epsilon}\right)^2\nonumber\\
  &=1+\lambda\left(\psi_1-\frac2{r_h}\right)\rho
  +\mathcal{O}(\lambda^2).
  \label{eq:post-channel-ratio}
\end{align}

Define the dimensionless fractional distance and the first conversion coefficient by
\begin{equation}
  \begin{gathered}
    \zeta\equiv\frac{\lambda\rho}{r_h}=\frac{r-r_h}{r_h},\\
    \mathcal D_h\equiv r_h\psi_1-2.
  \end{gathered}
\end{equation}
The variable $\zeta$ measures the fractional areal-radius displacement, and $\mathcal D_h$ is the linear coefficient of the exact geometric conversion. All $\mathcal{O}(\lambda^n)$ remainders in this result and in the black-hole applications are taken as $\lambda\to0$ at fixed $\rho$, for which $\zeta=\mathcal{O}(\lambda)$. Rewriting the expansion as $\mathcal R_s^{(h)}(\rho)/\mathcal R_s^{(h)}(0) =1+\mathcal D_h\zeta+\mathcal{O}(\lambda^2)$ is not a uniform error estimate at fixed nonzero $\zeta$.

To convert to the parent orthonormal frame, choose the tangent-space orientation so that the longitudinal and transverse components correspond to $(01)$ and $(23)$. Since $\lambda=\epsilon^2$, the exact decomposition in Eq.~\eqref{eq:sss-exact-spin-decomposition} gives
\begin{align}
  \Omega_{\rm phys}^{01}(\rho,\epsilon)
  &=\varsigma\epsilon^3\bar s_\parallel(\rho)
  +\mathcal{O}(\varsigma\epsilon^5)+\mathcal{O}(\varsigma^2),\\
  \Omega_{\rm phys}^{23}(\rho,\epsilon)
  &=\varsigma\epsilon\bar s_\perp(\rho)
  +\mathcal{O}(\varsigma\epsilon^3)+\mathcal{O}(\varsigma^2).
\end{align}
These are the parent-frame boost-generator and rotation-generator components, respectively. The $\epsilon$ dependence of the barred amplitudes is implicit. Their ratio is
\begin{equation}
  \mathcal R_{\Omega}(\rho,\epsilon)
  \equiv\frac{\Omega_{\rm phys}^{23}}{\Omega_{\rm phys}^{01}}
  =\epsilon^{-2}\bar{\mathcal R}_s(\rho).
  \label{eq:parent-lorentz-channel-ratio}
\end{equation}
This ratio generally has no finite $\epsilon\to0$ limit when $\bar{\mathcal R}_s$ is nonzero. It compares a boost-generator component with a rotation-generator component. It is not a ratio of two spatial rest-frame polarizations or, without a detector prescription, an operational observable. The radial normalization removes this $\rho$-independent weight.
\begin{equation}
  \begin{gathered}
    \frac{\mathcal R_{\Omega}(\rho,\epsilon)}
  {\mathcal R_{\Omega}(0,\epsilon)}
  =\frac{\bar{\mathcal R}_s(\rho)}{\bar{\mathcal R}_s(0)}
  =1,\\
    \epsilon\ \text{fixed}.
  \label{eq:normalized-parent-lorentz-channel-ratio}
  \end{gathered}
\end{equation}
$\mathcal D_h$ describes the fixed-to-local contracted basis conversion. The normalized radial change of the parent Lorentz-sector ratio at fixed contraction parameter vanishes exactly within the ideal radial spherical sector.

The coefficient of the curvature spin term enters at $\mathcal{O}(\lambda^2)$. For the orthonormal pair $(u^i,z^i)$, Eq.~\eqref{eq:sss-first-spin-force} gives
\begin{equation}
  \begin{gathered}
    u^i{\cal F}_i=0,\\
    z^i{\cal F}_i =\frac{\lambda^2}{2}{\cal K}_h s_\parallel +\mathcal{O}(\lambda^3),
  \end{gathered}
\end{equation}
where $\mathcal K_h=f_2+3f_1\psi_1$. The source is orthogonal to $u^i$ and therefore does not enter the comoving energy equation. Its longitudinal momentum projection depends only on the contracted boost-sector amplitude $s_\parallel$. Local contracted normalization changes it only at $\mathcal{O}(\lambda^3)$. The sign comparisons below keep the sign of $s_\parallel$ fixed. Keeping the parent component $\Omega_{\rm phys}^{01}$ finite and nonzero as $\epsilon\to0$ would require $s_\parallel=\mathcal{O}(\epsilon^{-3})$, outside the finite-current sector chosen in Eq.~\eqref{eq:sc-spin-current-series}. Under the Frenkel condition, Eq.~\eqref{eq:sss-frenkel-removes-longitudinal-spin} gives $s_\parallel=0$, so this curvature source vanishes.

In four dimensions, define $\ell_s\equiv s_\parallel/w_0$ and the linear-spin coefficient $\mathfrak f_s\equiv z^i{\cal F}_i/w_0$. Their dimensions are $L$ and $L^{-1}$, respectively, and
\begin{equation}
  \mathfrak f_s =\frac{\lambda^2}{2}{\cal K}_h\ell_s+\mathcal{O}(\lambda^3).
\end{equation}

Define the dimensionless curvature-source coefficient
\begin{equation}
  \mathcal J_h \equiv\frac{r_h{\cal K}_h}{f_1} =r_h\left(\frac{f_2}{f_1}+3\psi_1\right) =\frac{N_h r_h{\cal K}_h}{2\kappa}.
\end{equation}
Using $\kappa=N_hf_1/2$, the source-per-enthalpy scale becomes
\begin{align}
  \frac{N_h\,\mathfrak f_s^{\rm phys}}{\kappa}
  &=\varsigma\lambda^2\frac{\ell_s}{r_h}\mathcal J_h
  +\mathcal{O}(\varsigma\lambda^3)+\mathcal{O}(\varsigma^2),\\
  \mathfrak f_s^{\rm phys}
  &=\varsigma\mathfrak f_s+\mathcal{O}(\varsigma^2).
\end{align}
The normalized radial ratio in Eq.~\eqref{eq:normalized-parent-lorentz-channel-ratio} equals one within the ideal radial spherical sector. The coefficient $\mathcal D_h$ records only the fixed-to-local basis conversion. By contrast, $\mathcal J_h$ is a dimensionless form of the invariant longitudinal curvature scalar that enters $\mathcal G_{\rm sp}$ and $\mathcal A_{\rm sp}$. The response is not specified by $\mathcal J_h$ alone: it also depends on the signed spin flux, enthalpy, equation of state, and distance from the sonic point through Eqs.~\eqref{eq:stationary-spin-response-factor}--\eqref{eq:fixed-radius-linear-spin-response}.

\section{Black-hole applications and comparison with Carroll limits}
\label{sec:black-hole-applications}

The conversion and source coefficients depend on local derivatives of the metric. We evaluate them for the four-dimensional Einstein--Maxwell--dilaton (EMD) family, whose redshift function is generally nonconstant in areal-radius gauge. Let $a\geq0$ be the dilaton coupling and $R$ the standard radial coordinate. The metric is
\begin{subequations}\label{eq:emd-standard-metric}
  \begin{align}
    ds^2&=-A(R)dt^2+\frac{dR^2}{A(R)}+C(R)d\Omega_2^2,\\ A(R)&=\left(1-\frac{R_+}{R}\right) \left(1-\frac{R_-}{R}\right)^{(1-a^2)/(1+a^2)},\\ C(R)&=R^2 \left(1-\frac{R_-}{R}\right)^{2a^2/(1+a^2)}.
  \end{align}
\end{subequations}

This is the Gibbons--Maeda charged dilaton family, and the $a=1$ case is the Garfinkle--Horowitz--Strominger solution \cite{Gibbons:1987ps,Garfinkle:1990qj}. With the asymptotic dilaton set to zero, the geometrized ADM mass $M$ and electric charge $Q$ satisfy
\begin{equation}
  \begin{gathered}
    2M=R_++\frac{1-a^2}{1+a^2}R_-,\\
    Q^2=\frac{R_+R_-}{1+a^2}.
  \end{gathered}
\end{equation}

We restrict to the regular nonextremal branch $R_+>R_-\geq0$. For $a\neq0$, $R=R_-$ is a curvature singularity rather than a regular inner horizon, and $R_-<R_+$ keeps it inside the event horizon \cite{Gibbons:1987ps,Garfinkle:1990qj}.

The metric reduces to Schwarzschild when $R_-=0$ \cite{Schwarzschild:1916uq,Gibbons:1987ps}. For $a=0$, $C(R)=R^2$ and
\begin{equation}
  A(R)=1-\frac{2M}{R}+\frac{Q^2}{R^2},
\end{equation}
which is the Reissner--Nordstr\"om metric \cite{Reissner:1916,Nordstrom:1918,Gibbons:1987ps}. For $a=1$, $A(R)=1-2M/R$ and $C(R)=R(R-Q^2/M)$, giving the Einstein-frame Garfinkle--Horowitz--Strominger metric \cite{Garfinkle:1990qj}.

To convert Eq.~\eqref{eq:emd-standard-metric} to the areal gauge of Eq.~\eqref{eq:sss-static-metric}, define
\begin{subequations}\label{eq:emd-auxiliary-functions}
  \begin{align}
  &\begin{gathered}
      \Xi\equiv1+a^2,\\
      \gamma_a\equiv\frac{a^2}{\Xi},
    \end{gathered}\\
  &\begin{gathered}
      B(R)\equiv1-\frac{R_-}{R},\\
      H(R)\equiv1-\frac{R_-}{\Xi R}.
    \end{gathered}
\end{align}
\end{subequations}

The areal radius is $r\equiv\sqrt{C(R)}=R B(R)^{\gamma_a}$, and
\begin{equation}
  \frac{dr}{dR}=B^{\gamma_a-1}H.
\end{equation}

Matching the radial term to $dr^2/F(r)$ and using $g_{tt}=-e^{2\psi(r)}F(r)$ gives
\begin{subequations}\label{eq:emd-areal-gauge-functions}
  \begin{align}
    F(r(R)) &= \left(1-\frac{R_+}{R}\right)\frac{H(R)^2}{B(R)},\\ e^{\psi(r(R))} &=\frac{B(R)^{1/\Xi}}{H(R)}.
  \end{align}
\end{subequations}

The nonzero $\psi(r)$ comes entirely from the change to areal radius. Treating $R$ as the areal radius would miss this term.

Define
\begin{equation}
  \begin{gathered}
    \sigma\equiv\frac{R_-}{R_+},\\
    0\leq\sigma<1.
  \end{gathered}
\end{equation}

The event-horizon areal radius is $r_h=R_+(1-\sigma)^{\gamma_a}$. Evaluating $N_h=e^{\psi_h}$, $f_1=F'(r_h)$, $f_2=F''(r_h)$, and $\psi_1=\psi'(r_h)$ gives

\begin{align}
  &\begin{gathered}
      N_h=\frac{\Xi(1-\sigma)^{1/\Xi}}{\Xi-\sigma},\\
      f_1=\frac{\Xi-\sigma}{\Xi r_h},
    \end{gathered}\\
  &\begin{gathered}
      \psi_1=\frac{a^2\sigma^2} {r_h(\Xi-\sigma)^2},\\
      \kappa=\frac{1}{2R_+} (1-\sigma)^{(1-a^2)/\Xi}.
    \end{gathered}
\end{align}
\begin{equation}
  \begin{split}
    f_2={}&\frac{-2\Xi^2+6\Xi\sigma}{\Xi(\Xi-\sigma)r_h^2}\\
    &-\frac{(4+3a^2)\sigma^2}{\Xi(\Xi-\sigma)r_h^2}.
  \end{split}
\end{equation}
For fixed $a\geq0$ and $0\leq\sigma<1$, all denominators are nonzero and $\kappa>0$.

These quantities give the component-conversion and curvature-source coefficients
\begin{subequations}\label{eq:emd-spin-coefficients}
  \begin{align}
    \mathcal D_h &=-2+\frac{a^2\sigma^2}{(\Xi-\sigma)^2},\\ r_h^2\mathcal K_h &=-2+\frac{4\sigma}{\Xi},\\ \mathcal J_h &=-2\,\frac{\Xi-2\sigma}{\Xi-\sigma}.
  \end{align}
\end{subequations}

Here $\mathcal D_h$ is the fixed-basis conversion coefficient, while $\mathcal K_h$ and $\mathcal J_h$ describe the curvature source. With $\zeta=(r-r_h)/r_h$ and $\ell_s=s_\parallel/w_0$,

\begin{align}
  \frac{\mathcal R_s^{(h)}(\rho)}{\mathcal R_s^{(h)}(0)}
  &=1+ \left[-2+\frac{a^2\sigma^2}{(\Xi-\sigma)^2}\right]\zeta +\mathcal{O}(\lambda^2),\\
  \frac{\bar{\mathcal R}_s(\rho)}{\bar{\mathcal R}_s(0)}
  &=1,\\
  \frac{N_h\mathfrak f_s^{\rm phys}}{\kappa}
  &=-2\varsigma\lambda^2\frac{\ell_s}{r_h}
  \frac{\Xi-2\sigma}{\Xi-\sigma}
  +\mathcal{O}(\varsigma\lambda^3)+\mathcal{O}(\varsigma^2),\\
  \mathcal G_{\rm sp}^{\rm EMD}(\rho)
  &=1+\varsigma\lambda^2\frac{N_hj_\parallel\rho}
  {\Phi_{E,h}r_h^2}\left(1-\frac{2\sigma}{\Xi}\right)
  +\mathcal{O}(\varsigma\lambda^3)+\mathcal{O}(\varsigma^2).
\end{align}

The first line is a fixed-basis conversion, the second is the locally normalized ratio, and the last two lines give the force scale and the integrated matter-flow response. From Eq.~\eqref{eq:normalized-parent-lorentz-channel-ratio}, the second line is also the normalized radial change of the parent-frame Lorentz-sector ratio at fixed $\epsilon$. The response factor becomes unity when this curvature scalar vanishes or when the longitudinal spin flux is zero.

At fixed coupling, differentiation with respect to $\sigma$ gives

\begin{align}
  \partial_\sigma\mathcal D_h &=\frac{2a^2\Xi\sigma}{(\Xi-\sigma)^3},\\ \partial_\sigma\mathcal J_h &=\frac{2\Xi}{(\Xi-\sigma)^2},\\ \partial_\sigma(r_h^2\mathcal K_h)&=\frac4{\Xi}.
\end{align}

Both $\mathcal J_h$ and $r_h^2\mathcal K_h$ increase with $\sigma$. For $a>0$, $\mathcal D_h$ also increases, but this only changes fixed-basis components. The curvature source vanishes at
\begin{equation}
  \sigma_{\mathcal K}=\frac{1+a^2}{2}. \label{eq:emd-force-zero}
\end{equation}

This zero lies in $0<\sigma<1$ only for $0\leq a<1$. Its locus approaches the singular corner $(a,\sigma)=(1,1)$ as $a\to1^-$, but the corner is outside the regular branch and is not a zero of $\mathcal K_h$ on the exact $a=1$ family. For $a>1$, the algebraic locus lies outside the branch. At the interior zero, both $\mathcal G_{\rm sp}-1$ and $\mathcal A_{\rm sp}$ vanish at $\mathcal{O}(\varsigma\lambda^2)$. Across it, their signed changes also depend on the orientation of $j_\parallel$ or $s_\parallel$. In the Frenkel limit, $s_\parallel=j_\parallel=0$ and the response is absent. A zero of $\mathcal D_h$ does not define an independent physical locus because $\mathcal D_h$ cancels from Eq.~\eqref{eq:local-channel-ratio}.

Setting $\sigma=0$ gives Schwarzschild. The dilaton and Maxwell fields then vanish for any $a$. With $R_+=2M$, the Schwarzschild metric \cite{Schwarzschild:1916uq} gives
\begin{equation}
  \begin{gathered}
    r_h=2M,\\
    f_1=\frac1{r_h},\\
    f_2=-\frac2{r_h^2},\\
    \kappa=\frac1{2r_h},
  \end{gathered}
\end{equation}
which gives
\begin{equation}
  \begin{gathered}
    \mathcal D_h=-2,\\
    \mathcal K_h=-\frac2{r_h^2},\\
    \mathcal J_h=-2.
  \end{gathered}
\end{equation}

The physical signed momentum source is
$z^i\mathcal F_i^{\rm phys}
=-\varsigma\lambda^2s_\parallel/r_h^2
+\mathcal{O}(\varsigma\lambda^3)+\mathcal{O}(\varsigma^2)$.

The Reissner--Nordstr\"om (RN) limit follows by setting $a=0$ \cite{Reissner:1916,Nordstrom:1918}. For geometrized mass $M>0$ and charge $|Q|<M$, define

\begin{align}
R_\pm&=M\pm\sqrt{M^2-Q^2},\\
  &\begin{gathered}
      \delta\equiv\sqrt{1-\frac{Q^2}{M^2}},\\
      0<\delta\leq1.
    \end{gathered}
\end{align}

Then $\sigma=(1-\delta)/(1+\delta)$, $r_h=R_+$, and $\psi_1=0$. The horizon derivatives are

\begin{align}
  &\begin{gathered}
      f_1=\frac{R_+-R_-}{R_+^2},\\
      f_2=\frac{2(2R_--R_+)}{R_+^3},
    \end{gathered}\\
\kappa&=\frac{R_+-R_-}{2R_+^2}.
\end{align}
These expressions give
\begin{equation}
  \begin{gathered}
    \mathcal D_h=-2,\\
    \mathcal J_h=\frac1\delta-3,\\
    R_+^2\mathcal K_h =\frac{2(1-3\delta)}{1+\delta}.
  \end{gathered}
\end{equation}

The curvature coefficient vanishes at
\begin{equation}
  \begin{gathered}
    \delta=\frac13,\\
    \frac{|Q|}{M}=\frac{2\sqrt2}{3},
  \end{gathered}
\end{equation}

which is the $a=0$ case of Eq.~\eqref{eq:emd-force-zero}. Since $\psi_1=0$, RN changes the curvature-source coefficient but keeps $\mathcal D_h=-2$. At $Q=0$, all RN coefficients reduce to their Schwarzschild values.

This charge is not a new curvature threshold. In RN geodesic deviation, the horizon radial tidal eigenvalue changes from stretching to compression at the same value \cite{Crispino:2016rnm}. Equation~\eqref{eq:sss-longitudinal-curvature-scalar} explains the equality: both results are zeros of the same longitudinal curvature scalar. The new statement here is that the linear-spin canonical response in Eqs.~\eqref{eq:stationary-spin-response-factor}, \eqref{eq:fixed-radius-linear-spin-response}, and \eqref{eq:invariant-spin-response-scalar} vanishes at that established tidal zero.

The RN extremal point is not a uniform limit of this nonextremal expansion. To test the finite-distance radial truncation separately from the fixed-$\rho$ string-Carroll limit, compare the quadratic and linear terms in the ordinary Taylor expansion of $F$ while holding $\zeta$ fixed.
\begin{equation}
  \begin{gathered}
    \frac{\tfrac12f_2(\lambda\rho)^2} {f_1\lambda\rho} =\frac{\zeta}{2}\mathcal J_h,\\
    \zeta=\frac{\lambda\rho}{R_+}. \label{eq:rn-expansion-validity}
  \end{gathered}
\end{equation}

Equation~\eqref{eq:rn-expansion-validity} is a truncation estimate, not the same asymptotic statement as the fixed-$\rho$ $\mathcal{O}(\lambda^n)$ expansion. The truncation requires $|\zeta\mathcal J_h|/2\ll1$. Since $\mathcal J_h=\delta^{-1}-3$, this condition fails at fixed nonzero $\zeta$ as $\delta\to0^+$. A direct analysis of the RN limit finds that the second-order term in ingoing coordinates is needed to recover the radial dependence of the extremal throat. In static coordinates, the radial sector requires contributions from all orders \cite{Shinde:2026ern}. A regular treatment of the extremal endpoint requires a separate scaling \cite{Kunduri:2013gce}.

The $a=1$ branch is the GMGHS black hole \cite{Garfinkle:1990qj}. For nonzero charge, its redshift gradient differs from those of the Schwarzschild and RN solutions. For this branch, $R_+=2M$, $R_-=Q^2/M$, and $\sigma=Q^2/(2M^2)$ with $0\leq\sigma<1$. The horizon radius and surface gravity are
\begin{equation}
  \begin{gathered}
    r_h=2M\sqrt{1-\sigma},\\
    \kappa=\frac1{4M},
  \end{gathered}
\end{equation}
and the frame-conversion and curvature-source coefficients are

\begin{align}
  \mathcal D_h &=-2+\frac{\sigma^2}{(2-\sigma)^2},\\ \mathcal K_h &=-\frac{1}{2M^2},\\ \mathcal J_h &=-4\frac{1-\sigma}{2-\sigma},\\ r_h^2\mathcal K_h &=-2(1-\sigma).
\end{align}

For $0<\sigma<1$, $\psi_1\neq0$, so the fixed-basis conversion differs from the RN case. The locally normalized contracted ratio still obeys Eq.~\eqref{eq:local-channel-ratio}, and the corresponding normalized parent-frame Lorentz-sector ratio obeys Eq.~\eqref{eq:normalized-parent-lorentz-channel-ratio}. On the exact $a=1$ branch, $\mathcal K_h=-1/(2M^2)$ remains negative and nonzero throughout the regular domain and in its formal $\sigma\to1^-$ limit. By contrast, $r_h^2\mathcal K_h$ and $\mathcal J_h$ tend to zero because $r_h\to0$. The endpoint is singular and lies outside the assumed regular areal-radius class \cite{Garfinkle:1990qj}. The uncharged limit is Schwarzschild.

The metric blocks agree with earlier two-longitudinal string-Carroll expansions \cite{Bagchi:2023cfp,Bagchi:2024rje}. The coefficient $\mathcal D_h$ changes component frames, while $\mathcal J_h$ is a dimensionless form of this curvature scalar entering the canonical force. Neither is a string-sigma-model observable. The stationary response additionally contains the spin flux, enthalpy, equation of state, and sonic denominator through Eqs.~\eqref{eq:stationary-response-dressed-solution} and \eqref{eq:fixed-radius-linear-spin-response}.

The rank-one theory of Ref.~\cite{Shukla:2026chs} has one preferred null direction and the ideal spin current $\mathcal S^{\mu\nu\lambda}=k^\mu\mathfrak s^{\nu\lambda}$. For an $n$-dimensional longitudinal null space, the antisymmetric bivectors with both legs in that null space form $\Lambda^2(\ker h_p)$, which has dimension $n(n-1)/2$. The longitudinal spin counts are
\begin{equation}
  \begin{aligned}
    \dim\ker h_p=1&\ \longrightarrow\
    \dim\Lambda^2(\ker h_p)=\binom{1}{2}=0, \\
    \dim\ker h_p=2&\ \longrightarrow\
    \dim\Lambda^2(\ker h_p)=\binom{2}{2}=1.
  \end{aligned}
\end{equation}
The one-dimensional Carroll longitudinal null space therefore cannot support an antisymmetric spin component with two longitudinal legs. The two-dimensional string-Carroll longitudinal null space supports one such component, proportional to the oriented bivector $e_0\wedge e_1$. This component is the longitudinal Lorentz-boost spin amplitude used in the spherical reduction. The null-space conventions agree with standard Carroll and string-Carroll geometry \cite{Duval:2014uoa,Freidel:2022bai,Bagchi:2023cfp,Bergshoeff:2023ogz}.

The longitudinal contracted spin amplitude and the ratio $\bar{\mathcal R}_s=\bar s_\perp/\bar s_\parallel$ have no rank-one counterpart. Independently, reducing the dimension of the longitudinal null space from two to one removes the second longitudinal momentum projection.

The parent equations also differ. Ref.~\cite{Shukla:2026chs} contracts separately conserved ideal currents, while we contract the canonical stress identity with its Riemann spin source. Reducing the longitudinal null space to one direction gives a kinematic relation, not a full dynamical equivalence \cite{Hongo:2021ona,Gallegos:2022jow,Chiarini:2024cuv}.

The Schwarzschild, RN, and EMD examples separate the basis conversion from the physical curvature response. In each case, the normalized radial change of the parent-frame Lorentz-sector ratio at fixed $\epsilon$ equals one within the ideal radial spherical sector, while the absolute component ratio retains the $\epsilon^{-2}$ weight in Eq.~\eqref{eq:parent-lorentz-channel-ratio}. The spin-induced flow depends on $f_2+3f_1\psi_1$, not only on the surface gravity or horizon radius. Equations~\eqref{eq:stationary-response-dressed-solution}, \eqref{eq:fixed-radius-linear-spin-response}, and \eqref{eq:invariant-spin-response-scalar} give the complete matched coefficient at $\mathcal{O}(\varsigma\lambda^2)$ within the linear-spin ideal-stress closure before the sonic point; Eq.~\eqref{eq:stationary-invariant-killing-flux} provides its pseudo-gauge-independent flux check.

\section{Discussion and conclusions}
\label{sec:discussion-conclusions}

We contracted the canonical stress--spin Ward system in a rank-two string-Carroll geometry. For a finite mixed stress tensor and $S^{\lambda\mu\nu}=\mathcal{O}(\varsigma\epsilon^p)$, the first generic Ward order separates into passive ($p>0$), balanced ($p=0$), and singular ($p<0$) sectors. The balanced sector is the unique analytic finite-current choice for which the curvature source and induced stress response enter at the same generic order. Accidental zeros can postpone this order, and fractional $p$ requires a nonanalytic expansion. Lowering the free stress index before contraction removes the apparent $\epsilon^{-4}$ curvature term. The two-dimensional longitudinal null space supplies a second longitudinal momentum equation and one longitudinal antisymmetric bivector, whereas a one-dimensional longitudinal null space permits none.

The constitutive model retains the longitudinal bivector through an electric/boost susceptibility $\chi_{\rm B}<0$ and treats the transverse rotation sector with $\chi_{\rm R}>0$. These are necessary low-momentum signs and avoid the one-susceptibility sign conflict; they do not establish thermodynamic completeness, nonlinear causality, or well-posedness \cite{Daher:2024sce,Abboud:2025psf}. The strict Frenkel limit $\chi_{\rm B}=0$ removes the longitudinal component and every spherical response proportional to it.

A smooth static spherical outer horizon with a regular areal radius and $F'(r_h)>0$ lies on the regular branch in ingoing coordinates. Locally normalized longitudinal and transverse bivectors are parallel transported by an ideal radial flow, so their amplitude ratio is advected exactly. The varying ratio in a horizon-fixed basis is only the conversion $(N_\epsilon/N_h)(r_h/r_\epsilon)^2$. It is not a transport coefficient or an operational polarization observable.

The independent parameter $\varsigma$ fixes the ordering of the two asymptotic expansions. We take $\partial_\varsigma|_0$ before the $\lambda$ expansion, and the ideal stress has no explicit term linear in the spin potential. Quadratic terms such as $\mu_{\rm s}\Omega$ are therefore consistently excluded rather than inadvertently placed ahead of the curvature response \cite{Florkowski:2024gtr,Drogosz:2024sff,Armas:2026tif}. Within this stated closure, the complete matched result is the coefficient at $\mathcal{O}(\varsigma\lambda^2)$. The general barotrope is reduced by Eq.~\eqref{eq:stationary-general-response-solution}; the affine family has the elementary parametric solution in Eq.~\eqref{eq:stationary-response-dressed-solution} and the explicit fixed-radius coefficients in Eq.~\eqref{eq:fixed-radius-linear-spin-response}. For dust, $\Delta_{\rm sp}a_\parallel=\varsigma\lambda^2\mathcal K_h\ell_s/2+\mathcal{O}(\varsigma\lambda^3)+\mathcal{O}(\varsigma^2)$. The proof following Eq.~\eqref{eq:pseudogauge-killing-superpotential} establishes the pseudo-gauge invariance of the stationary closed-sphere Killing flux under its stated regularity assumptions.

In RN, $\mathcal K_h$ vanishes at $|Q|/M=2\sqrt2/3$. That value is the established radial-tidal inversion at the event horizon \cite{Crispino:2016rnm}; the result obtained here is that the spin-induced stationary response is governed by the same curvature scalar and vanishes there. In EMD, the corresponding zero is $\sigma=(1+a^2)/2$ and belongs to the regular branch only for $0\leq a<1$ \cite{Gibbons:1987ps,Garfinkle:1990qj}. On the exact $a=1$ GMGHS branch, $\mathcal K_h=-1/(2M^2)$ remains nonzero. The sign of the flow change also depends on the orientation of the longitudinal spin flux.

Compared with the 2026 rank-one formulation of Ref.~\cite{Shukla:2026chs}, our equations extend that setup in the respects relevant to the present problem. We use the coupled canonical Ward system, retain curvature, allow a rank-two degenerate longitudinal subspace, and obtain an analytic spherical response. The arbitrary-null-space count $\dim\Lambda^2(\ker h_p)=n(n-1)/2$ is kinematic. We make no claim about a complete constitutive dynamics for every $n$. All $\mathcal{O}(\lambda^n)$ statements use the fixed-$\rho$ contraction. The fixed-$\zeta$ RN test shows that this expansion is not uniform near extremality. Rotation, extremality, quadratic spin feedback, derivative spin transport, dissipation, physical torsion, backreaction, the sonic layer, and global accretion remain separate problems \cite{Kunduri:2013gce,Shinde:2026ern}.

\begin{acknowledgments}
  N.S.L. acknowledges institutional support from the University of Santo Tomas. R. P. would like to acknowledge networking support of the COST Action CA21106 - COSMIC WISPers in the Dark Universe: Theory, astrophysics and experiments (CosmicWISPers), the COST Action CA22113 - Fundamental challenges in theoretical physics (THEORY-CHALLENGES), the COST Action CA21136 - Addressing observational tensions in cosmology with systematics and fundamental physics (CosmoVerse), the COST Action CA23130 - Bridging high and low energies in search of quantum gravity (BridgeQG), and the COST Action CA23115 - Relativistic Quantum Information (RQI) funded by COST (European Cooperation in Science and Technology). R. P. would also like to acknowledge the funding support of SCOAP3.
\end{acknowledgments}

\section*{Data Availability statement}
No numerical or observational records were generated or analyzed for this study.

\appendix

\section{Rank-two connection, curvature, and projection algebra}
\label{app:rank-two-algebra}

The rank-two metric decomposition in Eq.~\eqref{eq:rank-two-metric} follows from the longitudinal and transverse frames. Define the two mixed projectors by
\begin{equation}
  \begin{gathered}
    P_{\parallel}{}^\mu{}_\nu=V^{\mu\rho}V_{\rho\nu},\\
    P_{\perp}{}^\mu{}_\nu=\Pi^{\mu\rho}\Pi_{\rho\nu}.
  \end{gathered}
\end{equation}
Here $V_{\mu\nu}$ and $V^{\mu\nu}$ are the longitudinal covariant and contravariant tensors, while $\Pi_{\mu\nu}$ and $\Pi^{\mu\nu}$ are their transverse counterparts. The inverse-frame relations give
\begin{subequations}\label{eq:app-a-projector-algebra}
  \begin{align}
  &\begin{gathered}
      V_{\mu\rho}\Pi^{\rho\nu}=0,\\
      \Pi_{\mu\rho}V^{\rho\nu}=0,
    \end{gathered}\\
  &\begin{gathered}
      P_{\parallel}^2=P_{\parallel},\\
      P_{\perp}^2=P_{\perp},
    \end{gathered}\\
  &\begin{gathered}
      P_{\parallel}P_{\perp}=0,\\
      P_{\parallel}+P_{\perp}=\mathbf 1.
    \end{gathered}
\end{align}
\end{subequations}
The traces are $2$ and $d-2$, respectively. These relations give
\begin{equation}
  \left(\epsilon^2V_{\mu\rho}+\Pi_{\mu\rho}\right)
  \left(\epsilon^{-2}V^{\rho\nu}+\Pi^{\rho\nu}\right) =\delta_\mu{}^\nu.
\end{equation}
This verifies the inverse metric used in the contraction without expanding in $\epsilon$.

The trace of the singular connection coefficient can be fixed before computing individual components. The physical coframe has two columns scaled by $\epsilon$, so, for spacetime-independent $\epsilon$,
\begin{subequations}\label{eq:app-a-trace-connection}
  \begin{align}
    \det\widehat E_\mu{}^I
    &=\epsilon^2\det E_\mu{}^I,\\
    \Gamma^\mu{}_{\mu\nu}
    &=\partial_\nu\ln\sqrt{-g}
    =\partial_\nu\ln|\det E_\mu{}^I|.
  \end{align}
\end{subequations}
The connection trace has no $\epsilon^{-2}$ term. In the notation of Eq.~\eqref{eq:sc-connection-series}, this gives $\Gamma^{(-2)\mu}{}_{\mu\nu}=0$.

For the full connection algebra, define

\begin{align}
  V_{\lambda\mu\nu}
  &=\frac12\left(\partial_\mu V_{\lambda\nu}
    +\partial_\nu V_{\lambda\mu}
  -\partial_\lambda V_{\mu\nu}\right),\\
  \Pi_{\lambda\mu\nu}
  &=\frac12\left(\partial_\mu \Pi_{\lambda\nu}
    +\partial_\nu \Pi_{\lambda\mu}
  -\partial_\lambda \Pi_{\mu\nu}\right).
\end{align}
At fixed $\epsilon$, substitution of the exact metric and its inverse into the Christoffel formula gives
\begin{align}
  \Gamma^\rho{}_{\mu\nu}
  &=\frac12\left(\epsilon^{-2}V^{\rho\lambda}
  +\Pi^{\rho\lambda}\right)
  \left(2\epsilon^2V_{\lambda\mu\nu}
  +2\Pi_{\lambda\mu\nu}\right)\nonumber\\
  &=\epsilon^{-2}{\cal A}^\rho{}_{\mu\nu}
  +{\cal C}^\rho{}_{\mu\nu}
  +\epsilon^2{\cal B}^\rho{}_{\mu\nu},
\end{align}
where

\begin{align}
  {\cal A}^\rho{}_{\mu\nu}
  &=V^{\rho\lambda}\Pi_{\lambda\mu\nu},\\
  {\cal C}^\rho{}_{\mu\nu}
  &=V^{\rho\lambda}V_{\lambda\mu\nu}
  +\Pi^{\rho\lambda}\Pi_{\lambda\mu\nu},\\
  {\cal B}^\rho{}_{\mu\nu}
  &=\Pi^{\rho\lambda}V_{\lambda\mu\nu}.
\end{align}
The even frame expansion makes each of ${\cal A}$, ${\cal C}$, and ${\cal B}$ analytic in $\epsilon^2$. Writing, for example, ${\cal A}=\sum_{j\geq0}\epsilon^{2j}{\cal A}_{(2j)}$, the first three connection coefficients are

\begin{align}
  \Gamma^{(-2)}&={\cal A}_{(0)},\\
  \Gamma^{(0)}&={\cal A}_{(2)}+{\cal C}_{(0)},\\
  \Gamma^{(2)}&={\cal A}_{(4)}+{\cal C}_{(2)}+{\cal B}_{(0)}.
\end{align}

The curvature orders follow without choosing coordinates. For two connection-type coefficients $X^\rho{}_{\mu\nu}$ and $Y^\rho{}_{\mu\nu}$, define the bilinear product
\begin{equation}
  (X\star Y)^\rho{}_{\sigma\mu\nu}
  \equiv2X^\rho{}_{[\mu|\lambda|}
  Y^\lambda{}_{\nu]\sigma}.
\end{equation}
Inserting $\Gamma=\epsilon^{-2}\Gamma^{(-2)}+\Gamma^{(0)} +\epsilon^2\Gamma^{(2)}+\mathcal{O}(\epsilon^4)$ into the Riemann convention of Eq.~\eqref{eq:riemann-convention} gives

\begin{align}
  R^{(-4)}
  &=\Gamma^{(-2)}\star\Gamma^{(-2)},\\
  R^{(-2)\rho}{}_{\sigma\mu\nu}
  &=2\partial_{[\mu}\Gamma^{(-2)\rho}{}_{\nu]\sigma}
  +(\Gamma^{(-2)}\star\Gamma^{(0)})^\rho{}_{\sigma\mu\nu}\\
  &\quad
  +(\Gamma^{(0)}\star\Gamma^{(-2)})^\rho{}_{\sigma\mu\nu},\nonumber\\
  R^{(0)\rho}{}_{\sigma\mu\nu}
  &=R^\rho{}_{\sigma\mu\nu}[\Gamma^{(0)}]
  +(\Gamma^{(-2)}\star\Gamma^{(2)})^\rho{}_{\sigma\mu\nu}\\
  &\quad
  +(\Gamma^{(2)}\star\Gamma^{(-2)})^\rho{}_{\sigma\mu\nu}. \nonumber
\end{align}
The notation $R[\Gamma^{(0)}]$ means the ordinary Riemann expression formed only from $\Gamma^{(0)}$.

The apparent $\epsilon^{-4}$ curvature is absent from the covariantly lowered tensor. From ${\cal A}^\rho{}_{\mu\nu}=V^{\rho\lambda} \Pi_{\lambda\mu\nu}$, the upper index of every $\Gamma^{(-2)\rho}{}_{\mu\nu}$ lies in the longitudinal image. Since the leading covariant metric $h_{\mu\nu}$ is transverse,
\begin{equation}
  \begin{gathered}
    h_{\alpha\rho}\Gamma^{(-2)\rho}{}_{\mu\nu}=0,\\
    h_{\alpha\rho}R^{(-4)\rho}{}_{\sigma\mu\nu}=0.
  \end{gathered}
\end{equation}
For $g_{\mu\nu}=h_{\mu\nu}+\epsilon^2k_{\mu\nu} +\epsilon^4\ell_{\mu\nu}+\mathcal{O}(\epsilon^6)$, index lowering gives

\begin{align}
  R_{\nu\alpha\beta\gamma}
  &=\epsilon^{-2}{\mathfrak R}^{(-2)}_{\nu\alpha\beta\gamma}
  +{\mathfrak R}^{(0)}_{\nu\alpha\beta\gamma}
  +\mathcal{O}(\epsilon^2),\\
  {\mathfrak R}^{(-2)}_{\nu\alpha\beta\gamma}
  &=h_{\nu\rho}R^{(-2)\rho}{}_{\alpha\beta\gamma}
  +k_{\nu\rho}R^{(-4)\rho}{}_{\alpha\beta\gamma},\\
  {\mathfrak R}^{(0)}_{\nu\alpha\beta\gamma}
  &=h_{\nu\rho}R^{(0)\rho}{}_{\alpha\beta\gamma}
  +k_{\nu\rho}R^{(-2)\rho}{}_{\alpha\beta\gamma}\\
  &\quad
  +\ell_{\nu\rho}R^{(-4)\rho}{}_{\alpha\beta\gamma}. \nonumber
\end{align}
The first possible order of the curvature spin source is $\mathcal{O}(\epsilon^{-2})$. With $S^{\alpha\beta\gamma}=s^{\alpha\beta\gamma} +\epsilon^2s_{(2)}^{\alpha\beta\gamma}+\mathcal{O}(\epsilon^4)$, its first two coefficients are
\begin{subequations}\label{eq:app-a-force-coefficients}
  \begin{align}
    {\cal F}^{(-2)}_\nu
    &=-\frac12{\mathfrak R}^{(-2)}_{\nu\alpha\beta\gamma}
    s^{\alpha\beta\gamma},\\
    {\cal F}^{(0)}_\nu
    &=-\frac12{\mathfrak R}^{(0)}_{\nu\alpha\beta\gamma}
    s^{\alpha\beta\gamma}
    -\frac12{\mathfrak R}^{(-2)}_{\nu\alpha\beta\gamma}
    s_{(2)}^{\alpha\beta\gamma}.
  \end{align}
\end{subequations}
Here ${\cal F}_\nu\equiv -R_{\nu\alpha\beta\gamma}S^{\alpha\beta\gamma}/2$.

The singular pieces of the two covariant divergences have the same order. For $T^\mu{}_\nu=\tau^\mu{}_\nu+\mathcal{O}(\epsilon^2)$ and $S^{\lambda\mu\nu}=s^{\lambda\mu\nu}+\mathcal{O}(\epsilon^2)$, expansion of the connection terms gives
\begin{subequations}\label{eq:app-a-singular-divergences}
  \begin{align}
    [\nabla_\mu T^\mu{}_\nu]_{-2}
    &=-\Gamma^{(-2)\rho}{}_{\mu\nu}\tau^\mu{}_\rho,\\
    [\nabla_\lambda S^{\lambda\mu\nu}]_{-2}
    &=\Gamma^{(-2)\mu}{}_{\lambda\rho}s^{\lambda\rho\nu}
    +\Gamma^{(-2)\nu}{}_{\lambda\rho}s^{\lambda\mu\rho}.
  \end{align}
\end{subequations}
The subscript $-2$ denotes the coefficient of $\epsilon^{-2}$. Terms containing the connection trace are absent by Eq.~\eqref{eq:app-a-trace-connection}. Equations \eqref{eq:app-a-force-coefficients} and \eqref{eq:app-a-singular-divergences} show that the finite current normalization used in the main contraction balances the curvature and divergence terms at the same power of $\epsilon$.

On the regular branch, the relations $u^\mu=v^{\mu\nu}\vartheta_\nu$, $v^{\mu\nu}\vartheta_\mu\vartheta_\nu=-1$, and $\bar\nabla_\rho v^{\mu\nu}=0$ give
\begin{align}
  \vartheta_\nu\bar\nabla_\mu u^\nu
  &=\vartheta_\nu v^{\nu\rho}
  \bar\nabla_\mu\vartheta_\rho\nonumber\\
  &=\frac12\bar\nabla_\mu
  \left(v^{\rho\sigma}\vartheta_\rho\vartheta_\sigma\right)=0.
\end{align}
No additional term proportional to $\vartheta_\nu\bar\nabla_\mu u^\nu$ survives in the longitudinal energy equation of Eq.~\eqref{eq:sc-longitudinal-equations}.

In block form, the parent mixed Lorentz parameter is $\beta_{\rm L}^A{}_a=-\epsilon\Lambda^A{}_a$. Define the $2\times(d-2)$ matrix $\boldsymbol{\Lambda}$ by $(\boldsymbol{\Lambda})^A{}_a=\Lambda^A{}_a$. In the strict limit, the mixed transformation acts on a two-index projected tangent tensor through
\begin{equation}
  \begin{gathered}
    M=
  \begin{pmatrix}0\boldsymbol{\Lambda}\\
    00
  \end{pmatrix},\\
    \delta_{\rm B}X=MX+XM^{\rm T},
  \end{gathered}
\end{equation}
where the first block is longitudinal and the second is transverse. For the spin current, this matrix acts only on the antisymmetric pair at fixed unprojected spacetime flux index. Writing that pair as $S=\bigl(
  \begin{smallmatrix}S_{LL}&S_{LT}\\ -S_{LT}^{\rm T}&S_{TT}
\end{smallmatrix}\bigr)$, this gives
\begin{equation}
  \delta_{\rm B}S=
  \begin{pmatrix}
    -\boldsymbol{\Lambda} S_{LT}^{\rm T}+S_{LT}\boldsymbol{\Lambda}^{\rm T}
    &\boldsymbol{\Lambda} S_{TT}\\
    S_{TT}\boldsymbol{\Lambda}^{\rm T}&0
  \end{pmatrix}.
\end{equation}
The transverse block is invariant, the mixed block is sourced only by the transverse block, and the longitudinal block is sourced only by mixed components. The same matrix rule without imposing antisymmetry gives the stress-tensor transformations in Eq.~\eqref{eq:stress-block-boost}. The intrinsic tensors $v^{\mu\nu}$ and $h_{\mu\nu}$ are unchanged because the inverse longitudinal frame and transverse coframe are boost invariant. The variations of the complementary projectors cancel in their sum, $\delta_{\rm B}(P_\parallel+P_\perp)=0$, and preserve the idempotence and orthogonality relations in Eq.~\eqref{eq:app-a-projector-algebra}.

\section{Post-string-Carroll expansion and analytic checks}
\label{app:analytic-checks}

The near-horizon expansion follows from the regular ingoing metric. Starting from Eq.~\eqref{eq:sss-static-metric}, the coordinate differential $dt=dv-e^{-\psi(r)}dr/F(r)$ gives
\begin{align}
  -e^{2\psi}Fdt^2+\frac{dr^2}{F}
  &=-e^{2\psi}Fdv^2+2e^\psi dv\,dr.
\end{align}
The $dr^2$ terms cancel. Set $r=r_h+\lambda\rho$, where $\lambda$ is dimensionless and $\rho$ has length dimension. With $N_h\equiv e^{\psi_h}$, $f_1\equiv F'(r_h)$, $f_2\equiv F''(r_h)$, and $\psi_1\equiv\psi'(r_h)$, the required Taylor products are
\begin{subequations}\label{eq:app-b-taylor-products}
  \begin{align}
    F&=f_1\lambda\rho
    +\frac12f_2\lambda^2\rho^2+\mathcal{O}(\lambda^3),\\
    e^\psi&=N_h\left(1+\psi_1\lambda\rho\right)
    +\mathcal{O}(\lambda^2),\\
    e^{2\psi}F
    &=N_h^2\left[f_1\lambda\rho
      +\left(\frac{f_2}{2}+2f_1\psi_1\right)
    \lambda^2\rho^2\right]+\mathcal{O}(\lambda^3).
  \end{align}
\end{subequations}
Since $dr=\lambda d\rho$, the nonzero metric components through $\mathcal{O}(\lambda^2)$ are

\begin{align}
  g_{vv}&=-N_h^2f_1\lambda\rho
  -N_h^2\left(\frac{f_2}{2}+2f_1\psi_1\right)
  \lambda^2\rho^2+\mathcal{O}(\lambda^3),\\
  g_{v\rho}&=N_h\lambda
  +N_h\psi_1\lambda^2\rho+\mathcal{O}(\lambda^3),\\
  g_{\theta\theta}&=r_h^2+2r_h\lambda\rho
  +\lambda^2\rho^2,\\
  g_{\phi\phi}&=g_{\theta\theta}\sin^2\theta.
\end{align}
With $\lambda=\epsilon^2$, the terms proportional to $\lambda$ and $\lambda^2$ are the $\epsilon^2$ and $\epsilon^4$ coefficients, respectively.

Writing $N(r)\equiv e^{\psi(r)}$, the inverse of the unexpanded $(v,\rho)$ block has components
\begin{subequations}\label{eq:app-b-inverse-base}
  \begin{align}
  &\begin{gathered}
      g^{vv}=0,\\
      g^{v\rho}=\frac{1}{\lambda N},
    \end{gathered}\\
g^{\rho\rho}&=\frac{F}{\lambda^2}.
\end{align}
\end{subequations}
Substitution of Eqs.~\eqref{eq:app-b-taylor-products} and \eqref{eq:app-b-inverse-base} in the Christoffel formula gives the finite leading base connection

\begin{align}
  &\begin{gathered}
      \bar\Gamma^v{}_{vv}=\kappa,\\
      \bar\Gamma^\rho{}_{v\rho}=-\kappa,
    \end{gathered}\\
\bar\Gamma^\rho{}_{vv}&=2\kappa^2\rho,
\end{align}
where $\kappa=N_hf_1/2$. No $\lambda^{-1}$ coefficient occurs. The coefficient of the first correction, $\Gamma^\mu{}_{\nu\sigma}=\bar\Gamma^\mu{}_{\nu\sigma} +\lambda\Gamma^{(2)\mu}{}_{\nu\sigma}+\mathcal{O}(\lambda^2)$, is

\begin{align}
\Gamma^{(2)v}{}_{vv}
  &=\frac{N_h}{2}(f_2+3f_1\psi_1)\rho,\\
\Gamma^{(2)\rho}{}_{v\rho}
  &=-\Gamma^{(2)v}{}_{vv},\\
\Gamma^{(2)\rho}{}_{vv}
  &=\frac{N_h^2f_1}{4}
  (3f_2+8f_1\psi_1)\rho^2,\\
\Gamma^{(2)\rho}{}_{\rho\rho}
  &=\psi_1,\\
  &\begin{gathered}
      \Gamma^{(2)\theta}{}_{\rho\theta}
  =\frac1{r_h},\\
      \Gamma^{(2)\phi}{}_{\rho\phi}=\frac1{r_h}.
    \end{gathered}
\end{align}
The correction trace $C_i\equiv\Gamma^{(2)\mu}{}_{\mu i}$ satisfies
\begin{equation}
  \begin{gathered}
    C_v=0,\\
    C_\rho=\psi_1+\frac{2}{r_h}.
  \end{gathered}
\end{equation}
These coefficients enter the first post-string-Carroll stress and spin equations.

The curvature combination in the spin force can be computed from the unexpanded metric without expanding the connection. Direct evaluation gives
\begin{equation}
  R_{vrvr}=\frac{e^{2\psi}}{2}
  \left[F''+3\psi'F'
  +2F\left(\psi''+(\psi')^2\right)\right].
\end{equation}
At the horizon, $F(r_h)=0$, so define

\begin{align}
  {\cal K}_h&\equiv f_2+3f_1\psi_1,\\
  R_{vrvr}\big|_{r_h}&=\frac{N_h^2}{2}{\cal K}_h.
\end{align}
Because each covariant $\rho$ index contributes $dr/d\rho=\lambda$,
\begin{equation}
  R_{v\rho v\rho}
  =\frac{\lambda^2N_h^2}{2}{\cal K}_h
  +\mathcal{O}(\lambda^3).
\end{equation}
For the invariant longitudinal spin bivector $\varepsilon_\parallel^{v\rho}=N_h^{-1}$, contraction with $\Sigma^{\mu\nu}=s_\parallel \varepsilon_\parallel^{\mu\nu}+ s_\perp\varepsilon_\perp^{\mu\nu}$ gives
\begin{subequations}\label{eq:app-b-force-components}
  \begin{align}
    {\cal F}_v
    &=-\frac{\lambda^2N_h}{2}{\cal K}_h
    u^\rho s_\parallel+\mathcal{O}(\lambda^3),\\
    {\cal F}_\rho
    &=\frac{\lambda^2N_h}{2}{\cal K}_h
    u^v s_\parallel+\mathcal{O}(\lambda^3).
  \end{align}
\end{subequations}
The warped-product curvature has no component with a longitudinal radial-flow index and an antisymmetric pair entirely on the two-sphere, so $s_\perp$ is absent. For the normalized longitudinal vectors $u^i$ and $z^i$, $z^\rho u^v-z^v u^\rho=N_h^{-1}$. Equation \eqref{eq:app-b-force-components} then gives

\begin{align}
  u^i{\cal F}_i&=0,\\
  z^i{\cal F}_i
  &=\frac{\lambda^2}{2}{\cal K}_h s_\parallel
  +\mathcal{O}(\lambda^3).
\end{align}

For an algebraic check of the stationary solution, use the affine equation of state in Eq.~\eqref{eq:affine-barotrope}. With $R_w=w_0/w_h$, the density factor is
\begin{equation}
  \begin{gathered}
    \frac{{\cal N}}{{\cal N}_h}=R_w^{1/(1+\alpha)},\\
    {\cal N}_h={\cal N}({\cal E}_h)=1.
  \label{eq:app-b-density-factor-linear}
  \end{gathered}
\end{equation}
Let $\Delta\eta=\eta-\eta_h$. Ratios of the conserved density and matter energy fluxes to their horizon values give
\begin{subequations}\label{eq:app-b-flux-ratios}
  \begin{align}
    R_w^{1/(1+\alpha)}e^{-\Delta\eta}(1-y)&=1,\\
    R_w e^{-2\Delta\eta}(1-y^2)&=1,
  \end{align}
\end{subequations}
where $y=N_h\kappa\rho e^{2\eta}$. Eliminating $\Delta\eta$ gives
\begin{subequations}\label{eq:app-b-stationary-elimination}
  \begin{align}
    R_w&=\left(\frac{1+y}{1-y}\right)^{(1+\alpha)/(1-\alpha)},\\
    e^{\Delta\eta}
    &=(1+y)^{1/(1-\alpha)}
    (1-y)^{-\alpha/(1-\alpha)}.
  \end{align}
\end{subequations}
Since $x=N_h\kappa\rho e^{2\eta_h}$, $y/x=e^{2\Delta\eta}$ and hence
\begin{equation}
  x=y(1-y)^{2\alpha/(1-\alpha)}
  (1+y)^{-2/(1-\alpha)}.
  \label{eq:app-b-parametric-map}
\end{equation}
Its logarithmic derivative is
\begin{align}
  \frac{d\ln x}{dy}
  &=\frac{D_0(y)}{(1-\alpha)y(1-y^2)},\\
  D_0(y)&=(1-y)^2-\alpha(1+y)^2.
\end{align}
The first positive zero for $0<\alpha<1$ is $y=(1-\sqrt\alpha)/(1+\sqrt\alpha)$, which is also the zero of Eq.~\eqref{eq:stationary-sonic-denominator}. The second root, $(1+\sqrt\alpha)/(1-\sqrt\alpha)$, lies beyond $y=1$.

At linear order in $\varsigma$, the density flux is unchanged while the matter energy-flux ratio is $\mathcal G_{\rm sp}=1+\varsigma g$. Thus
\begin{align}
  R_w^{1/(1+\alpha)}e^{-\Delta\eta}(1-y)&=1,\\
  R_we^{-2\Delta\eta}(1-y^2)&=\mathcal G_{\rm sp}.
\end{align}
Elimination gives
\begin{equation}
  R_w=\left[\frac{1+y}{(1-y)\mathcal G_{\rm sp}}
  \right]^{(1+\alpha)/(1-\alpha)},
\end{equation}
and the remaining relations in Eq.~\eqref{eq:stationary-response-dressed-solution}. At fixed $x$, write $y=y_0+\varsigma y_{[1]}+\mathcal{O}(\varsigma^2)$. The first variation of the logarithm of the parametric map is
\begin{equation}
  0=\frac{D_0(y_0)}{(1-\alpha)y_0(1-y_0^2)}y_{[1]}
  +\frac{2g}{1-\alpha},
\end{equation}
which gives the first line of Eq.~\eqref{eq:fixed-radius-linear-spin-response}. The other two independent variations are
\begin{align}
  (\ln R_w)_{[1]}
  &=\frac{1+\alpha}{1-\alpha}
  \left(\frac{2y_{[1]}}{1-y_0^2}-g\right),\\
  \eta_{[1]}
  &=\frac{y_{[1]}}{1-\alpha}
  \left(\frac{1}{1+y_0}+\frac{\alpha}{1-y_0}\right)
  -\frac{g}{1-\alpha}.
\end{align}
Substitution of $y_{[1]}$ yields the remaining rational expressions in Eq.~\eqref{eq:fixed-radius-linear-spin-response}.

A differential check follows directly from the stationary energy and momentum equations. At a fixed local state,
\begin{equation}
  \begin{pmatrix}
    u^\rho&w_0z^\rho\\
    \alpha z^\rho&w_0u^\rho
  \end{pmatrix}
  \begin{pmatrix}{\cal E}_0'\\ \eta'
  \end{pmatrix}
  =
  \begin{pmatrix}
    -w_0\kappa z^v\\
    z^i{\cal F}_i-w_0\kappa u^v
  \end{pmatrix}.
\end{equation}
The determinant is $w_0\Delta_{\rm s}$. Away from the sonic point, the direct source-induced changes of the derivatives are
\begin{equation}
  \begin{gathered}
    \delta_{\cal F}{\cal E}_0'
  =-\frac{z^\rho}{\Delta_{\rm s}}z^i{\cal F}_i,\\
    \delta_{\cal F}\eta'
  =\frac{u^\rho}{w_0\Delta_{\rm s}}z^i{\cal F}_i.
  \end{gathered}
\end{equation}
This matrix check shows explicitly why the curvature force is not equal to acceleration for $\alpha>0$ and why the integrated solution must include both pressure and rapidity corrections.

For pressureless matter, setting $\alpha=0$ and $\Pi=0$ in Eqs.~\eqref{eq:app-b-stationary-elimination} and \eqref{eq:app-b-parametric-map} gives
\begin{align}
  &\begin{gathered}
      \frac{{\cal E}_0}{{\cal E}_h}=\frac{1+y}{1-y},\\
      e^{\Delta\eta}=1+y,\\
      x=\frac{y}{(1+y)^2}.
    \end{gathered}
\end{align}
The density relation in Eq.~\eqref{eq:app-b-density-factor-linear} gives the same factor for $s_X/s_{Xh}$. Together, these expressions reproduce Eq.~\eqref{eq:stationary-dust-limit} without a separate radial integral.

The fixed-basis component conversion does not require the velocity correction. Divide the stationary correction fluxes by $j_X=u^\rho s_X$, where $X$ labels the longitudinal or transverse channel. Their difference is
\begin{align}
  \frac{J_{\perp(2)}^\rho}{j_\perp}
  -\frac{J_{\parallel(2)}^\rho}{j_\parallel}
  &=C_0+\left(\psi_1-\frac{2}{r_h}\right)\rho,
\end{align}
where $C_0$ is the difference of the two horizon integration constants. The common term $u_{(2)}^\rho/u^\rho$ cancels when the left-hand side is converted to the fractional fixed-basis coefficients. Normalization at $\rho=0$ removes $C_0$, leaving the coefficient $\psi_1-2/r_h$ in Eq.~\eqref{eq:post-channel-ratio}. The same coefficient is produced by the relative normalization of the two bivectors in Eq.~\eqref{eq:sss-local-bivectors}. The result is
\begin{equation}
  \left[\delta_\lambda\ln\bar{\mathcal R}_s\right]_{\rho}
  =\left[\delta_\lambda\ln\mathcal R_s^{(h)}\right]_{\rho}
  -\left(\psi_1-\frac{2}{r_h}\right)\rho=0.
\end{equation}
Here $[\delta_\lambda(\cdot)]_\rho$ denotes the coefficient of the first-order variation in $\lambda$ at fixed $\rho$. This perturbative cancellation is the linear expansion of the exact equality in Eq.~\eqref{eq:local-channel-ratio}.

The Schwarzschild and Reissner--Nordstr\"om results follow by differentiating the metric functions. For Schwarzschild, $F=1-2M/r$, $\psi=0$, and $r_h=2M$ give

\begin{align}
  &\begin{gathered}
      f_1=\frac1{r_h},\\
      f_2=-\frac2{r_h^2},
    \end{gathered}\\
  &\begin{gathered}
      {\cal K}_h=-\frac2{r_h^2},\\
      {\cal J}_h\equiv\frac{r_h{\cal K}_h}{f_1}=-2.
    \end{gathered}
\end{align}
For Reissner--Nordstr\"om, $F=1-2M/r+Q^2/r^2$ and $\psi=0$. Introduce its two horizon radii $r_\pm$ through $r_++r_-=2M$ and $r_+r_-=Q^2$. Evaluation at $r_h=r_+$ gives

\begin{align}
  f_1&=\frac{r_+-r_-}{r_+^2},\\
  f_2&=\frac{2(2r_--r_+)}{r_+^3},\\
  {\cal J}_h
  &=\frac{2(2r_--r_+)}{r_+-r_-}.
\end{align}
Define $\delta\equiv\sqrt{1-Q^2/M^2}$ with $0<\delta\leq1$, so that $r_\pm=M(1\pm\delta)$. Then

\begin{align}
  {\cal J}_h&=\frac1\delta-3,\\
  r_+^2{\cal K}_h
  &=\frac{2(1-3\delta)}{1+\delta}.
\end{align}
At $Q=0$, $\delta=1$ and both expressions reduce to the Schwarzschild values. The curvature coefficient vanishes at $\delta=1/3$, which is equivalent to $|Q|/M=2\sqrt2/3$.

Let $\zeta\equiv\lambda\rho/r_+$ be the dimensionless distance from the outer RN horizon. This defines a finite-distance truncation test at fixed $\zeta$. It is separate from the $\lambda\to0$ contraction at fixed $\rho$ used for the $\mathcal{O}(\lambda^n)$ remainder notation. The ratio of the quadratic to the linear term in the Taylor series of $F$ is
\begin{equation}
  \frac{\tfrac12f_2(\lambda\rho)^2}
  {f_1\lambda\rho}
  =\frac{\zeta}{2}{\cal J}_h.
\end{equation}
The truncation requires $|\zeta{\cal J}_h|/2\ll1$. As $\delta\to0^+$, $r_+^2{\cal K}_h\to2$ but ${\cal J}_h\to+\infty$. The divergence comes from $f_1\to0$ and shows that the fixed-$\zeta$ nonextremal expansion is not uniform at the extremal endpoint.

For the EMD metric, use the definitions $\Xi=1+a^2$, $\gamma_a=a^2/\Xi$, $B=1-R_-/R$, and $H=1-R_-/({\Xi R})$ from Eq.~\eqref{eq:emd-auxiliary-functions}. Since $r=RB^{\gamma_a}$, the areal-radius chain rule is
\begin{equation}
  \frac{d}{dr}
  =\frac{B^{1/\Xi}}{H}\frac{d}{dR}.
\end{equation}
Applying this operator twice to the functions in Eq.~\eqref{eq:emd-areal-gauge-functions} and then setting $R=R_+$, $\sigma=R_-/R_+$, gives

\begin{align}
  &\begin{gathered}
      r_h\psi_1
  =\frac{a^2\sigma^2}{(\Xi-\sigma)^2},\\
      r_hf_1=\frac{\Xi-\sigma}{\Xi},
    \end{gathered}\\
r_h^2f_2
  &=\frac{-2\Xi^2+6\Xi\sigma
  -(4+3a^2)\sigma^2}
  {\Xi(\Xi-\sigma)}.
\end{align}
Substitution into $\mathcal D_h=r_h\psi_1-2$, $\mathcal K_h=f_2+3f_1\psi_1$, and $\mathcal J_h=r_h\mathcal K_h/f_1$ yields
\begin{subequations}\label{eq:app-b-emd-coefficient-check}
  \begin{align}
    \mathcal D_h
    &=-2+\frac{a^2\sigma^2}{(\Xi-\sigma)^2},\\
    r_h^2\mathcal K_h
    &=-2+\frac{4\sigma}{\Xi},\\
    \mathcal J_h
    &=-2\frac{\Xi-2\sigma}{\Xi-\sigma},
  \end{align}
\end{subequations}
which reproduces Eq.~\eqref{eq:emd-spin-coefficients}.

Removing charge and dilaton hair sets $\sigma=0$ and gives
\begin{equation}
  (\mathcal D_h,r_h^2\mathcal K_h,\mathcal J_h)
  =(-2,-2,-2),
  \label{eq:app-b-emd-schwarzschild-limit}
\end{equation}
the Schwarzschild result. For $a=0$, $\Xi=1$ and $\sigma=(1-\delta)/(1+\delta)$, so

\begin{align}
  &\begin{gathered}
      \mathcal D_h=-2,\\
      r_h^2\mathcal K_h
  =\frac{2(1-3\delta)}{1+\delta},
    \end{gathered}\\
\mathcal J_h&=\delta^{-1}-3,
\end{align}
which is the RN reduction. For $a=1$, $\Xi=2$, $R_+=2M$, and

\begin{align}
\mathcal D_h
  &=-2+\frac{\sigma^2}{(2-\sigma)^2},\\
  &\begin{gathered}
      \mathcal K_h=-\frac{1}{2M^2},\\
      r_h=2M\sqrt{1-\sigma},
    \end{gathered}\\
  &\begin{gathered}
      r_h^2\mathcal K_h=-2(1-\sigma),\\
      \mathcal J_h=-4\frac{1-\sigma}{2-\sigma},
    \end{gathered}
\end{align}
which is the GMGHS reduction. The neutral RN condition $\delta=1$ and the uncharged GMGHS condition $\sigma=0$ both return Eq.~\eqref{eq:app-b-emd-schwarzschild-limit}.

Equation~\eqref{eq:app-b-emd-coefficient-check} gives the curvature-source zero.
\begin{equation}
  \begin{gathered}
    \mathcal K_h=0\\
    \Longrightarrow\quad \sigma=\frac{1+a^2}{2},\\
    0\leq a<1.
  \end{gathered}
\end{equation}
For $a=0$, this locus gives $\sigma=1/2$, hence $\delta=(1-\sigma)/(1+\sigma)=1/3$ and $|Q|/M=2\sqrt2/3$. The domain restriction follows from requiring the source zero to lie inside $0<\sigma<1$. As $a\to1^-$, the zero locus approaches the excluded singular corner $(a,\sigma)=(1,1)$. On the exact $a=1$ family, $\mathcal K_h=-1/(2M^2)$ and does not vanish. A zero of $\mathcal D_h$ is not listed as a physical locus because $\mathcal D_h$ only converts the two contracted bivector bases and cancels from the normalized radial change of the parent-frame Lorentz-sector ratio in Eq.~\eqref{eq:normalized-parent-lorentz-channel-ratio}.

\bibliography{ref}

\end{document}